\documentclass[11pt,onecolumn]{article}
\usepackage{geometry,setspace}
\usepackage[font=small,format=plain,labelfont=bf,justification=raggedright,singlelinecheck=false]{caption}
\usepackage{hyperref,cite}
\usepackage{amsmath,amsfonts,amssymb,amsthm,upgreek,mathrsfs,bm}
\usepackage{graphicx,subfig}
\usepackage{tikz}
\usepackage{physics}
\usepackage[normalem]{ulem}

\hypersetup{colorlinks=true,linkcolor=blue,anchorcolor=blue,citecolor=blue,urlcolor=blue}
\allowdisplaybreaks[2]

\numberwithin{equation}{section}

\definecolor{darkgreen}{RGB}{40,150,60}

\title{Structure of Carrollian (conformal) superalgebra}
\author{Yu-fan Zheng$^{1}$\footnote{\href{zhengyufan@bimsa.cn}{zhengyufan@bimsa.cn}} and Bin Chen$^{2,3,4}$\footnote{Corresponding author: \href{chenbin1@nbu.edu.cn}{chenbin1@nbu.edu.cn}}}
\date{\today}

\begin{document}

\maketitle
\begin{center}
    {\it
        $^{1}$ Beijing Institute of Mathematical Sciences and Applications (BIMSA), Huaibei Town, Huairou District, Beijing 101408, China \\\vspace{2mm}
	$^{2}$Institute of Fundamental Physics and Quantum Technology,\\ \& School of Physical Science and Technology, \\ Ningbo University, Ningbo, Zhejiang 315211, China\\\vspace{2mm}
        $^3$School of Physics, Peking University, No.5 Yiheyuan Rd, Beijing 100871, P.~R.~China\\
	\vspace{2mm}
	$^{4}$Center for High Energy Physics, Peking University, No.5 Yiheyuan Rd, Beijing 100871, P.~R.~China\\
    }
    \vspace{10mm}
\end{center}

\begin{abstract}
    \vspace{5mm}
    \begin{spacing}{1.5}
        In this work, we investigate possible supersymmetric extensions of the Carrollian algebra and the Carrollian conformal algebra in both $d=4$ and $d=3$. For the super-Carrollian algebra in $d=4$, we identify multiple admissible structures, depending on the representations of the supercharges with respect to the Carrollian rotation. Some of these structures can be derived by taking the speed of light $c\to 0$ limit from super-Poincar\'e algebra, but others are completely novel. 
            In the conformal case, we derive nontrivial Carrollian superconformal algebras in dimensions $d=4$ and $d=3$. Among these, the superconformal algebra in $d=4$ and one of the algebras in $d=3$ exhibit isomorphisms to the super-Poincar\'e algebras in $d=5$ and $d=4$, respectively. Additionally, we identify a novel, nontrivial superconformal algebra in $d=3$ that is not isomorphic to any super-Poincar\'e algebra. 
        Remarkably, neither of these constructions requires R-symmetry to ensure the algebraic closure. Given that BMS$_4$ algebra constitutes the infinite-dimensional extension of the $d=3$ Carrollian conformal algebra, their supersymmetric extension gives rise to nontrivial superconformal Carrollian algebras. Specifically, we demonstrate the existence of a singlet super-BMS$_4$ algebra emerging from the extension of the $d=3$ Carrollian superconformal algebra, as well as a multiplet super-BMS$_4$ algebra that does not admit this methodology, as its finite-dimensional subalgebra incorporates supercharges with conformal dimension $\Delta=\pm\frac{3}{2}$.
    \end{spacing}
\end{abstract}

\setcounter{tocdepth}{2}
\tableofcontents

\section{Introduction}\label{sec:Introduction}
    
    The Carrollian contraction of Poincar\'e symmetry was proposed independently by L\'evy-Leblond \cite{Levy-Leblond:1965} and Sen Gupta \cite{Gupta:1966} in the 1960's, initially as a purely mathematical construction. In the ultra-relativistic $c\to0$ limit, the Minkowski metric becomes degenerate Carrollian metric:
    \begin{equation}
         \dd{s}^2 = 0 ~ \dd{t}^2 + \dd{x}^i \dd{x}^i
    \end{equation}
    Consequently, the symmetry of the space differs fundamentally from Poincar\'e symmetry. Essentially, the boost in Carrollian space only changes the time coordinate while keeping the space coordinates invariant:
    \begin{equation}
        t'=t-\vec{b}\cdot \vec {x}, \qquad \vec{x}~'=\vec{x}.
    \end{equation}
    In the Carrollian space, the light-cone closes, suggesting that information in the spacetime is frozen. At the time of its discovery, the significance of Carrollian symmetry was unrecognized, as there was no known physical system exhibiting such property. 
    \par
    
    It was not until recent decades that Carrollian physics emerged as a rapidly evolving research area, driven by its unique degenerate geometric structure and the fact that the Carrollian conformal symmetry is isomorphic to the asymptotic symmetry of flat spacetime \cite{Bondi:1962px, Sachs:1962wk, Sachs:1962zza, Barnich:2009se, Duval:2014uva, Duval:2014lpa, Duval:2014uoa}, which plays an indispensable role in flat space holography \cite{Barnich:2010eb, Bagchi:2010zz, Barnich:2012rz, Hartong:2015usd, Bagchi:2016bcd, Ciambelli:2019lap, Nguyen:2023vfz, Bekaert:2024itn, Chen:2023naw, Liu:2024nkc, Liu:2024llk} and celestial holography \cite{Donnay:2022aba, Bagchi:2022emh, Donnay:2022wvx, Adamo:2024mqn}. Remarkably, it was realized that the asymptotic symmetry at null infinity, the so-called Bondi--Metzner--Sachs (BMS) symmetry, is an infinite extension of Carrollian conformal symmetry. Moreover, the null boundary of a $d$-dimensional flat spacetime is naturally a $(d-1)$-dimensional Carrollian space. These facts strongly suggest that the boundary dual in flat-space holography is a Carrollian conformal theory. Furthermore, the Carrollian correlation function defined at null infinity plays a crucial role in connecting the bulk scattering matrix and the celestial correlation function. This connection provides a novel framework for investigating soft theorems and offers fresh perspectives on holographic correspondences in asymptotically flat spacetime. \par

    The Carrollian physics itself presents a rich physical structure, which has become an active area of research in recent years\cite{Basu:2018dub, Barducci:2018thr, Bagchi:2019clu, Bagchi:2019xfx, Chen:2020vvn, Banerjee:2020qjj, Chen:2021xkw, Fuentealba:2021xhn, Chen:2022jhx, Chen:2022cpx, Henneaux:2021yzg, Hao:2021urq, Rivera-Betancour:2022lkc, Hao:2022xhq, Yu:2022bcp, Baiguera:2022lsw, Bergshoeff:2022eog, Bergshoeff:2022qkx, Banerjee:2022ocj, Bagchi:2022eui, Bagchi:2022eav, Saha:2022gjw, Chen:2023pqf, Fuentealba:2023hzq, deBoer:2023fnj, Islam:2023rnc, Chen:2024voz, Chen:2025gaz, Tropper:2024evi, Tropper:2024kxy}. Among these developments, \cite{Chen:2021xkw} provided an explicit discussion on the structure of Carrollian conformal algebra and the corresponding representations in $d=4$ (with the case of $d=3$ addressed in its Appendix). Applying the theorem in \cite{Jakobsen:2011zz}, it was shown that the representations of the Carrollian rotation usually are reducible but not decomposable. More explicitly, the representation of the $4$D Carrollian rotation consists of $\mathfrak{so}(3)$ sectors connected by boost generators. One crucial property is that the boost maps sectors to sectors unidirectionally. These results not only inspired intense discussions of various kinds of Carrollian field theories (including the formulation of magnetic Carrollian theories), but also established the foundational framework of this work. \par

    Supersymmetry is a profound development in theoretical physics, providing a unified framework that incorporates both the bosonic and fermionic degrees of freedom. 
    The concept was first proposed by Golfand and Likhtman \cite{Golfand:1971iw}, who discovered the symmetry between bosons and fermions. Subsequently, supersymmetry was further developed by Volkov and Akulov \cite{Volkov:1973ix} formulating the Goldstino theory, as well as by Wess and Zumino \cite{Wess:1974tw} establishing the supersymmetric field theory in a mathematical consistent way. These works laid the foundation for modern supersymmetry research. Notably, supersymmetry presents a nontrivial extension of the Coleman--Mandula theorem for massive theory. It imposes strong restrictions on the physical Hilbert space in quantum field theory and plays the fundamental rule in supersymmetric field theory, supergravity\cite{deWit:2002vz}, superstring theory \cite{Green:1987sp, Polchinski:1998rr} as well as  holography\cite{Banks:2010tj}.  In the Carrollian context, several studies have explored fermionic theories in various dimensions \cite{Barducci:2018thr, Hao:2022xhq, Yu:2022bcp, Banerjee:2022ocj, Bagchi:2022owq, Bagchi:2022eui, Koutrolikos:2023evq, Kasikci:2023zdn, Bergshoeff:2023vfd, Zorba:2024jxb, Ekiz:2025hdn, Sharma:2025rug}. Besides, there are studies of super BMS$_3$ theories, which is equivalent to $2$D Carrollian superconformal theories \cite{Bagchi:2016yyf, Lodato:2016alv, Bagchi:2017cte, Bagchi:2018ryy, Chen:2023esw}. In particular, in the studies on super BMS$_3$ theories,  two kinds of supersymmetries in $d=2$, i.e. homogeneous and inhomogeneous supersymmetry, have been discovered. These two kinds of supersymmetries emerge from different rescalings of the supercharges in taking the $c\to 0$ limit, and they relate deeply with the fact that there exist electric and magnetic versions of Carrollian theories. Building on these insights, the aim of the present work is to deeply investigate supersymmetric extensions of Carrollian algebras in higher dimensions, especially in $d=4$ and $d=3$ to find the possible structure in the corresponding super-Lie algebra. \par
    
    In this work, we focus on investigating possible structures of supersymmetric extension\footnote{
        Throughout this paper, we do not consider general extended supersymmetry or general central charges.} of Carrollian algebra and Carrollian conformal algebra in $d=4$ and $d=3$.
    Based on \cite{Chen:2021xkw}, we consider the supercharges transforming in chain representations and adopt basic ansatz: $\{Q,Q\}\sim P, [P,Q]=0$. Through the analysis of the Jacobi identities, we find nontrivial super-Carrollian algebra with certain pattern for the $Q$ representations in $d=4$ case. Especially, the algebras presented in \eqref{eq:4dSuperCarrCase1} and \eqref{eq:4dSuperCarrCase2} can be obtained by taking the $c\to 0$ limit from super-Poincar\'e algebra with different rescaling of supercharge $Q$. However, we also discover additional nontrivial algebras that can not be derived by taking limits. Extending our analysis to the conformal case, we find that there are less allowed structures. By introducing special super-conformal generator $S$, and solving the Jacobi identities by using the basic ansatz \eqref{eq:4dsuperConCarrAnsatz}, we find only one nontrivial super-conformal Carrollian algebra with supercharges in \eqref{eq:4dSuperConCarrSolution3} and commutation relations in \eqref{eq:4dSuperConCarrCommutators}. Remarkably, this algebra is actually isomorphic to $5$D super-Poincar\'e algebra as expected. A significant distinction emerges when comparing with the $4$D superconformal algebra: while the R-symmetry is necessarily a part of $4$D superconformal algebra, the $4$D Carrollian superconformal algebra achieves closure without R-symmetry. From the perspective of the taking-limit procedure, the $U_R (1)$ R-symmetry either decouples from the algebra (in the sense that $R$ is absent from any commutators) or becomes a central charge, depending on specific rescaling of the $R$ charge with respect to powers of $c$. \par

    For the case of $d=3$, it is significantly challenging to systematically discuss the nonconformal superalgebra since the restriction from the Carrollian rotation is not enough. We therefore focus on the Carrollian superconformal algebra which is isomorphic to $4$D super-Poincar\'e algebra. Similarly, the algebra is also closed without R-symmetry. One special feature of $3$D Carrollian conformal algebra is that it can be infinitely extended to the BMS$_4$ algebra. In Section \ref{subsec:SingletSuperBMS4}, we construct such an extension to the singlet BMS$_4$ algebra. We discuss the possible Hermitian conjugation conditions, and we find that although all the symmetries are isomorphic to the BMS$_4$ algebra, the physical conjugation conditions for bulk $4$D flat theory, $3$D Carrollian theory, and $2$D Lorentzian celestial CFT are all different. In \cite{Bagchi:2016yyf, Lodato:2016alv, Bagchi:2017cte, Bagchi:2018ryy, Chen:2023esw}, the authors introduced two kinds of super BMS$_3$ algebra, i.e. the homogeneous version and inhomogeneous version. The singlet super-BMS$_4$ algebra is analogous to a homogeneous superalgebra. Moreover, in Section \ref{subsec:MultipletletSuperBMS4}, we discover two multiplet chiral super-BMS$_4$ algebras, which parallel to the inhomogeneous superalgebra. Remarkably, the multiplet algebras can not be found from Carrollian superconformal algebra since its finite subalgebra contains supercharges of conformal dimension $\Delta=\pm\frac{3}{2}$. \par

    This paper is organized as follows: we first review Carrollian conformal algebra and its representation in Section \ref{sec:CarrAlg}, as well as super-Poincar\'e algebra and superconformal algebra in Section \ref{sec:SUSY}, providing the preliminaries of this work. Readers who are familiar with these subjects can safely skip these sections. In Section \ref{sec:4dCarrSUSY}, we present the discussion on super-Carrollian algebra and Carrollian superconformal algebra in $d=4$, before turning to the case of $d=3$ in Section \ref{sec:3dCarrSUSY}. Especially, in Section \ref{subsec:SingletSuperBMS4} and \ref{subsec:MultipletletSuperBMS4}, we discuss two kinds of super-BMS$_4$ algebras respectively. Finally, we conclude this work in Section \ref{sec:Discussion}. Throughout this work, the Greek indices $\mu, \nu = 0, 1, ... , d-1$ denote the spacetime directions, while the Latin indices $i,j,=1,...,d-1$ refer only to the space directions. \par

\section{Carrollian and Carrollian conformal algebras}\label{sec:CarrAlg}

    A generic Carrollian manifold $\mathscr{C}$ is equipped with a degenerate metric $g_{\mu\nu}$ and a time-directing null vector $\zeta^\mu$, which satisfy the orthogonal condition $g_{\mu\nu}\zeta^\mu = 0$. The flat Carrollian space can be obtained by taking ultra-relativistic limit ($c\to 0$ limit) from Minkowski space. Under this limit, the metric and null vector take the following forms:
    \begin{equation}
        \dd{s}^2 =g_{\mu\nu} \dd{x}^\mu \dd{x}^\nu = 0 ~ \dd{t}^2 + \dd{x}^i \dd{x}^i, \quad \zeta = \partial_t .
    \end{equation}
    The Carrollian symmetry is defined as the finite dimensional isometries that preserve both degenerate metric $g_{\mu\nu}$ and the null vector $\zeta^\mu$\cite{Henneaux:2021yzg}. By solving the Killing equations
    \begin{equation}
        \mathcal{L}_{\hat \xi} g_{\mu\nu} = 0, \quad \mathcal{L}_{\hat \xi} \zeta^\mu = 0,
    \end{equation} 
    and restricting to the global transformations, we identify transformations constituting the finite dimensional Carrollian symmetry: the spacial rotations, the Carrollian boosts and the translations. Consequently, the Carrollian algebra is generated by $\{J^{ij},B^{i},P^{\mu}\}$ with the following commutation relations
    \begin{equation}
        \begin{aligned}
            &[J^{ij},J^{kl}]=\delta^{ik}J^{jl}-\delta^{il}J^{jk}+\delta^{jl}J^{ik}-\delta^{jk}J^{il}, \\
            &[J^{ij},P^k]=\delta^{ik}P^j-\delta^{jk}P^i,\quad [J^{ij},B^k]=\delta^{ik}B^j-\delta^{jk}B^i, \\
            &[B^i,P^j]=\delta^{ij}P^0.\\
        \end{aligned}
    \end{equation} 
    The rotation part of Carrollian algebra, consisting of spacial rotations $J^{ij}$ and Carrollian boosts $B^i$, is referred to as Carrollian rotations. \par
    
    The conformal extension of Carrollian symmetry is obtained by solving the conformal killing equations:
    \begin{equation}
        \mathcal{L}_\xi g_{\mu\nu} = \omega g_{\mu\nu}, \quad \mathcal{L}_\xi \zeta^\mu = \omega^{-\frac{1}{2}} \zeta^\mu,
    \end{equation}
    where $\omega(x^\mu)$ is an arbitrary function.\footnote{
        Generally, the Carrollian conformal Killing equations are given by $\mathcal{L}_\xi g_{\mu\nu} = \omega g_{\mu\nu}$ and $ \mathcal{L}_\xi \zeta^\mu = \omega^{-\frac{k}{2}} \zeta^\mu$, where the parameter $k$ is known as the level of the conformal Carrollian algebra\cite{Duval:2014lpa} and it determines the relative scaling between temporal and spatial directions. In this paper, we specifically focus on the case with $k=1$, where the temporal and spatial coordinates rescale isotropically.  }
    The solutions generating global transformations give the Carrollian conformal algebra. In addition to the generators of the Carrollian algebra, it includes dilation $D$ and special conformal transformations (SCTs) $K^\mu$, with the relations given by 
     \begin{equation}\label{eq:ConCarrAlg}
        \begin{aligned}
            &[D,P^\mu]=P^\mu, \quad [D,K^\mu]=-K^\mu, \quad [D,B^i]=[D,J^{ij}]=0, \\
            &[J^{ij},J^{kl}]=\delta^{ik}J^{jl}-\delta^{il}J^{jk}+\delta^{jl}J^{ik}-\delta^{jk}J^{il}, \\
            &[J^{ij},G^k]=\delta^{ik}G^j-\delta^{jk}G^i, \quad G\in\{P,K,B\}, \quad [J^{ij},P^0]=[J^{ij},K^0]=0,\\
            &[B^i,P^j]=\delta^{ij}P^0, \quad [B^i,K^j]=\delta^{ij}K^0, \quad [B^i,B^j]=[B^i,P^0]=[B^i,K^0]=0, \\
            &[K^0,P^0]=0, \quad [K^0,P^i]=-2B^i, \quad [K^i,P^0]=2B^i, \quad [K^i,P^j]=2\delta^{ij}D+2J^{ij}.
        \end{aligned}
    \end{equation}
    Remarkably, the $d$-dimensional Carrollian conformal algebra is isomorphic to $(d+1)$-dimensional Poincar\'e algebra $\mathfrak{cca}_d\equiv\mathfrak{iso}(1,d)\equiv\mathfrak{so}(1,d)\ltimes\mathbb{R}^{(d+1)}$. The dilation $D$, the spacial rotations $J^{ij}$, the spacial translation $P^i$, and the spacial SCT $K^i$ span the rotation part $\mathfrak{so}(1,d)=\mbox{span}\{D,J^{ij},P^i,K^i\}$, while the other part is given by Carrollian boosts $B^i$, the temporal translation $P^0$, and the temporal SCT $K^0$: $\mathbb{R}^{(d+1)}=\mbox{span}\{B^i,P^0,K^0\}$. This isomorphism plays a crucial role in the analysis of supersymmetric Carrollian conformal algebras. \par
    
    To facilitate the discussion of supersymmetric Carrollian algebras, it is necessary to review the finite dimensional representation of Carrollian rotations. The super-Carrollian generators $Q$ must belong to one of these representations, while the Carrollian superconformal generators $Q$ and $S$ must be in the same representation and possess eigenvalues $\pm\frac{1}{2}$ under the action of dilation $D$. Notably, the $d$-dimensional Carrollian rotation is isomorphic to $(d-1)$-dimensional Poincar\'e algebra $\mathfrak{iso}(1,d-1)$, which is not a semi-simple Lie algebra. The Jakobsen's theorem \cite{Jakobsen:2011zz} then tells us that the finite dimensional representation of Carrollian rotations is either trivial under boosts or reducible but not decomposable. A generic representation of Carrollian rotation algebra can be highly involved. In the following, we will focus on reviewing some typical structure of the representations for Carrollian rotations following \cite{Chen:2021xkw}. \par

    The Carrollian rotation in $4$ dimension includes an $\mathfrak{so}(3)$ subalgebra along with three boosts. Due to Jakobsen's theorem, the representations that are trivial under Carrollian boosts correspond to irreducible representations of $\mathfrak{so}(3)$ algebra. We denote the representation $(j)$ by its spin $j\in \mathbb{N}/2$. This simplest representation is referred to as singlet since there is only one $\mathfrak{so}(3)$ sub-sector. More complicated representations are called multiplet representations, which contain multiple $\mathfrak{so}(3)$ sub-sectors. Among these, the simplest case for multiplets are the so-called chain representations, denoted as 
    \begin{equation}
        (j_1)\to(j_2)\to\cdots\to(j_n).
    \end{equation}
    In this notation, $(j)$ represents a subsector transforming as $\mathfrak{so}(3)$ representation with spin $j$, and the arrow indicates the action of the boost generators. Noticing that the action of boost is unidirectional, meaning it only connects subsectors in one direction. These representations are called chain representations because they behave as a chain under the action of boosts. By the Jacobi identity, there are four kinds of chain representations:
    \begin{equation}
        \begin{aligned}
            (j)\to&(j),\\
            (0)\to(1)&\to(0),\\
            (j)\to(j+1)&\to(j+2)\to\cdots,\\
            (j+2)\to(j+&1)\to(j)\to\cdots,\\
        \end{aligned}
    \end{equation}
    with $j\in \mathbb{N}/2$. Additionally, there exist more complex structures known as net representations. However, they are highly complicated and are beyond the scope of this paper. \par

\begin{figure}[htpb]
        \centering
        \includegraphics[width=0.5\linewidth]{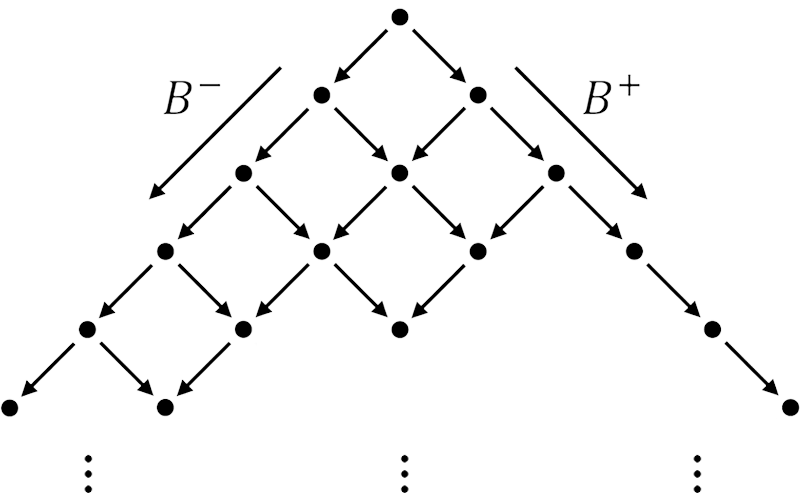}
        \caption{\centering An example of representation of $3$D Carrollian rotations. The dots are the states with certain spin, and the arrows represent the actions of boost generators $B^\pm$. }
        \label{fig:3dCarrollianRotationRepresentation}
    \end{figure}

    The representation of $3$D Carrollian rotation is less constrained compared to the $4$D case. The spacial rotation subalgebra is $\mathfrak{so}(2)=\mathfrak{u}(1)$, which implies that its irreducible representations are $1$-dimensional with arbitrary spin $j\in\mathbb{R}$. When the action of the boost generators is taken into account, a representation of $3$D Carrollian rotations is visualized as in Figure \ref{fig:3dCarrollianRotationRepresentation}. To obtain meaningful results, additional manual constraints must be imposed to specify the types of representations under consideration. Due to this difficulty, as well as the limited interest in $3$D super-Carrollian symmetries, we will focus more on Carrollian superconformal algebra in $3$D. \par

\section{Supersymmetry and superconformal symmetry}\label{sec:SUSY}

    This section provides an overview of supersymmetric algebra in Minkowski spacetime, establishing the basic concepts for the analysis of Carrollian extensions. \par
    
\subsection{Supersymmetric Poincar\'e algebra}\label{subsec:SuperPoincare}

    The Coleman--Mandula theorem and its supersymmetric generalization, the Haag--Łopuszański--Sohnius theorem, determine the structure of supersymmetry algebra, including the representation of supersymmetry generators under $\mathfrak{so}(1,3)$ and their (anti-)commutation relations. Here we give a brief review, while omitting the discussion on the cases with central charges and extended supersymmetry. \par
    
    The Coleman--Mandula theorem is a strong No-go theorem stating that the only symmetries for a $4$D massive Lorentzian theory with nontrivial scattering are the translations $P^\mu$ in $(\frac{1}{2}, \frac{1}{2})$ representation, the rotations $J^{\mu\nu}$ in $(1,0)\oplus(0,1)$ representation, and the internal symmetries in $(0,0)$ representation. Here $(j_L,j_R)$ denotes an irreducible representation of $\mathfrak{so}(1,3)=\mathfrak{so}_{L}(3)\otimes\mathfrak{so}_{R}(3)$, and $j_{L/R}$ is the spin of $\mathfrak{so}_{L/R}(3)$. The Haag--Łopuszański–-Sohnius theorem generalizes this theorem to the supersymmetric case and states that the supersymmetric generators $Q$ and $\bar{Q}$ are in $(\frac{1}{2}, 0)$ and $(0, \frac{1}{2})$ representations respectively. The tensor product of the representations $(\frac{1}{2}, 0)\otimes(0, \frac{1}{2})=(\frac{1}{2}, \frac{1}{2})$ indicates that the fermionic generators have $\{Q, \bar{Q}\}\propto P$, and the unitary condition further requires that this anti-commutation relation is not vanishing. Thus we know the commutation relation of $Q$ and $\bar{Q}$ are 
    \begin{equation}\label{eq:4dSuperPoincareAlgebra}
        \{Q_\alpha, \bar{Q}^{\dot{\beta}}\}= 2 g_{\mu\nu} (\sigma^\mu)_{\alpha}^{~~\dot{\beta}}P^\nu, \quad 
        \{Q_\alpha, Q_\beta\} = \{\bar{Q}^{\dot{\alpha}}, \bar{Q}^{\dot{\beta}}\} = 0.
    \end{equation}
    Here $\sigma^\mu$ are Dirac matrices. Moreover, considering the Jacobi identity of $P^\mu, P^\nu$ and $Q_\alpha$, we find that $[P,Q]=[P,\bar{Q}]=0$, and so we obtain the full super-Poincar\'e algebra $\mathfrak{iso}(1,3|1)$ containing $\{J^{\mu\nu}, P^\mu, Q, \bar{Q}\}$. \par

    Notice that these results depend crucially on the assumption that the rotation algebra is semi-simple $\mathfrak{so}(1,3)$ as well as the theory being unitary. As mentioned in Section \ref{sec:CarrAlg}, the Carrollian rotation algebra is not semi-simple. In addition, many of Carrollian theories, including the simplest scalar theories, are not unitary \cite{Chen:2024voz, Bekaert:2024itn}. Therefore, we can not use those beautiful No-go theorems in the Carrollian case. Instead, we consider $\{Q,Q\}\propto P$ and\footnote{
        Another reason to assume $[P,Q]=0$ is to keep the potential extension to conformal case. In the conformal case, $P$ has conformal dimension $1$ and $Q$ has conformal dimension $\frac{1}{2}$, and thus $[P,Q]$ should be of a conformal-dimension $\frac{3}{2}$ generator. However there is no such generator, so the commutator should vanish. } $[P,Q]=0$
    as the starting point in this work and discuss what kind of supersymmetric algebra we can obtain in the Carrollian case. \par

    Let us end this subsection by mentioning a non-intuitive fact that the anti-commutator of a fermionic generator with itself is not always vanishing. As a concrete example, notice that
    \begin{equation}
        \{Q_1,\bar{Q}^{\dot{1}}\}=-2(P^0 + P^3), \quad \{Q_1,Q_1\}=\{\bar{Q}^{\dot{1}},\bar{Q}^{\dot{1}}\}=0.
    \end{equation}
    We can define a fermionic generator $\tilde{Q}:=Q_1+\bar{Q}^{\dot{1}}$, which have nonvanishing anti-commutation relation with itself,
    \begin{equation}
        \{\tilde{Q},\tilde{Q}\}=-4(P^0 + P^3).
    \end{equation}
    So, the relations $\{Q_\alpha, Q_\beta\} = \{\bar{Q}^{\dot{\alpha}}, \bar{Q}^{\dot{\beta}}\} = 0$ are just a result but not a requirement of super-Poincar\'e algebra. Therefore, we will not assume $\{Q,Q\}=0$ in general. \par

\subsection{Supersymmetric conformal algebra}\label{subsec:SuperConformal}

    The Coleman--Mandula theorem does not apply to massless theories in which the conformal symmetry is allowed. The extra symmetry generators are dilation $D$ and SCTs $K^\mu$ in the conformal algebra. The generators of conformal algebra can be classified by their eigenvalues under $D$, which are referred to as conformal dimensions $\Delta$. The fermionic generators fit into this classification and have conformal dimensions $\pm\frac{1}{2}$. A consistent (anti-)commutator should have the same conformal dimension on both sides. Therefore, an anti-commutator between two dimension-$\frac{1}{2}$ generators must lead to dimension-$1$ generators. For instance, the anti-commutator between two supersymmetric generators takes the form: $\{Q,\bar{Q}\}\propto P$. In superconformal algebra, there exist two other fermionic generators $S, \bar{S}$ with dimension $-\frac{1}{2}$ and their anti-commutator gives a dimension $-1$ generator: $\{S,\bar{S}\}\propto K$. Here $S$ and $\bar{S}$ are in $(\frac{1}{2},0)$ and $(0,\frac{1}{2})$ representations under $\mathfrak{so}(1,3)$, respectively. The consistency in conformal dimension further indicates that $[P,S]\propto Q$, $[K,Q]\propto S$, as well as $\{Q,S\}\supset\# D+\#J$. \par

    \begin{table}[htpb]
        \renewcommand\arraystretch{1.8}
        \centering
        \caption{\centering Conformal dimension of superconformal generators.}
        \label{tab:SuperConCarrConDim}
        \begin{tabular}{c|c|c|c|c|c}
            \hline
            generator & $P^\mu$ & $Q, \bar{Q}$ & $D, J^{\mu\nu}, R$& $S, \bar{S}$ & $K^\mu$ \\
            \hline
            $\Delta$ & $1$ & $\frac{1}{2}$ & $0$ & $-\frac{1}{2}$ & $-1$ \\
            \hline
        \end{tabular}
    \end{table}

    Moreover, for the usual superconformal algebra in $d=4$, adding $Q$'s and $S$'s into the conformal algebra does not lead to a closed algebra. In order to have a closed algebra, it is necessary to include the generators of R-symmetry of supercharges $U_R(\mathcal{N})$. This is mandatory even for non-extended case, that is, for $\mathcal{N}=1$, the $U_R(1)$ generator $R$ is part of superconformal algebra. Thus the superconformal algebra can be fit into a superalgebra:
    \begin{equation}
        \mqty(
            D, J, P, K & Q, \bar{S} \\
            \bar{Q}, S & R, T^A \\
        ),
    \end{equation}
    where $R$ and $T^A$ are generators of $U_R(\mathcal{N})$ and $SU(\mathcal{N})$ in $U_R(\mathcal{N})=U_R(1)\otimes SU_R(\mathcal{N})$, respectively. Especially, the anti-commutation relations of the supercharges are roughly
    \begin{equation}
        \begin{aligned}
            &\{Q,S\}=\# D + \# J^{\mu\nu} + \# R + \# T^A, 
            &\{\bar{Q},\bar{S}\}=\# D + \# J^{\mu\nu} + \# R + \# T^A. \\
        \end{aligned}
    \end{equation}
    The explicit coefficients in the commutation relations can be found for example in \cite{Kovacs:1999fx, DHoker:2002nbb, Cordova:2016emh}. \par

    Later we will see that in Carrollian conformal algebra, the R-symmetries are not necessary in the supersymmetric extension. In the Carrollian case, we will take the dimensions of $Q$ and $S$ to be $\pm \frac{1}{2}$, as well as 
    \begin{equation}
        [P,S]\propto Q, \quad [K,Q]\propto S, \quad \{Q,S\}\supset\# D+\#J
    \end{equation}
    as the starting points to find closed Carrollian superconformal algebra, but not taking the existence of R-symmetry as a requirement. \par

\subsection{Superconformal symmetry in \texorpdfstring{$d=3$}{d=3}}

    The superconformal symmetry in $d=3$ is slightly simpler. Here the rotation algebra is $\mathfrak{so}(1,2)\simeq\mathfrak{so}(3)$, and thus the supercharge $Q$ is in spin-$\frac{1}{2}$ representation and satisfies
    \begin{equation}
        \{Q_{s_1},Q_{s_2}\} = \frac{1}{\sqrt{2}}\mqty(iP^1-P^2 & P^0\\P^0& -iP^1-P^2),
    \end{equation}
    such that the algebra is given by $\{J^{\mu\nu}, P^\mu, Q_{s}\}$.\par

    Extending to the conformal case, we see that it is enough just to introduce supercharge $Q$ and special supercharge $S$ with spin $\frac{1}{2}$ to find a closed super-Lie algebra, even without R-symmetry. This means that the $3$-dimensional superconformal algebra contains the generators $\{D,J^{\mu\nu}, P^\mu, K^\mu, Q_{s}, S_{s}\}$. Especially, the anti-commutation relations between the supercharges are
    \begin{equation}
        \begin{aligned}
            &\{Q_{s_1},Q_{s_2}\} = \frac{1}{\sqrt{2}}\mqty(iP^1-P^2 & P^0\\P^0& -iP^1-P^2), \\[0.5em]
            &\{S_{s_1},S_{s_2}\} = \frac{1}{\sqrt{2}}\mqty(iK^1-K^2 & K^0\\K^0& -iK^1-K^2), \\[0.5em]
            &\{Q_{s_1},S_{s_2}\} = \frac{1}{\sqrt{2}}\mqty(-iJ^{01}+J^{02} & -D-iJ^{12}\\D-iJ^{12}& -iJ^{01}-J^{02}).
        \end{aligned}
    \end{equation}
    \par

\section{Carrollian supersymmetry in \texorpdfstring{$d=4$}{d=4}}\label{sec:4dCarrSUSY}

    In this section, we focus on the $d=4$ case and try to find possible structures in the supersymmetric extension of Carrollian algebra and Carrollian conformal algebra. As introduced in Section \ref{sec:SUSY}, to find a super-Lie algebra extension of a given algebra we should first specify the assumptions. Here we make ansatz on the representations of supercharges under Carrollian rotations as introduced in Section \ref{sec:CarrAlg}, as well as (anti-)commutation relations among the generators of the algebra with undetermined coefficients. Then by solving the Jacobi identities for all combinations of the generators, we fix the coefficients to get the answer. \par
    
    In Section \ref{subsec:4dSuperCarr}, we shall show that there are multiple self-consistent super-Carrollian algebras, based on different representations of $Q$'s with respect to the Carrollian rotation. However, in Section \ref{subsec:4dSuperConCarr} we find that there is only one nontrivial Carrollian superconformal algebra, which is isomorphic to $5$D super-Poincar\'e algebra. This is what we expect: the $d$-dimensional Carrollian conformal algebra is the same as the $(d+1)$-dimensional Poincar\'e algebra, and so should their supersymmetric extension be. One surprising thing we find is that the Carrollian superconformal algebra does not require the R-symmetry, in contrast to the usual superconformal algebra. From perspective of taking the ultra-relativistic limit, the R-symmetry generator either decouples from $\{Q,S\}$ anti-commutator or becomes a central charge, depending on how we rescale the R symmetry in taking the limit. \par

\subsection{Carrollian supersymmetric algebra in \texorpdfstring{$d=4$}{d=4}}\label{subsec:4dSuperCarr}

    To find a self-contained super-Carrollian Lie algebra, we should make suitable ansatz on the supercharge $Q$. Firstly, we assume that the spins of supercharges under spacial rotation $\mathfrak{so}(3)$ are half-integers, and they are in singlet representations or chain representations. Hence there are four situations: singlets, a rank-2 chain with same spin (called it 2-chain for simplicity), a raising chain with spin increasing, and a lowering chain with spin decreasing:
    \begin{align}
        &\begin{aligned}
            &\mbox{singlet} \quad ~~ (j), \\
        \end{aligned}\label{eq:4dCarrSinglet}\\
        &\begin{aligned}
            &\mbox{2-chain} && (j)\to(j), \\
            &\mbox{raising} && (j)\to(j+1)\to(j+2)\to\cdots, \\
            &\mbox{lowering} && (j+2)\to(j+1)\to(j)\to\cdots ,&& j\in \mathbb{N}+\frac{1}{2}.\\
        \end{aligned}
    \end{align}
    As discussed in Section \ref{subsec:SuperPoincare}, the rest of ansatz is given by
    \begin{equation}
        [P^\mu,Q_a]=0, \quad \{Q_a,Q_b\}\propto P.
    \end{equation}
    Although the tensor products of $Q$ representations in the Carrollian case can generate same representation of $J$ or $B$, i.e. the spin-$1$ representations, we exclude $J$ and $B$ from the anti-commutation relations of $Q$: $\{Q,Q\}\not\supset J, B$. This restriction is motivated by the requirement that the super-Carrollian algebra should admit a natural extension to the Carrollian superconformal algebra. Crucially, $J$ and $B$ are of conformal dimension $1$, while $\{Q,Q\}$ is of dimension $0$, so we have $\{Q,Q\}\not\supset J, B$. \par
    
    For the reason explained in Section \ref{subsec:SuperPoincare}, we do not forbid nonvanishing anti-commutation relation of supercharge with itself, i.e. we assume $\{Q_a,Q_a\}\neq 0$ in general. On the other hand, one can always get a solution by setting $\{Q_a,Q_b\}=0$ for all supercharges. We call the solution with vanishing $\{Q,Q\}$ anti-commutation relations trivial, and call the solution with at least one nonvanishing $\{Q,Q\}$ anti-commutation relation nontrivial. \par

    In this work, we do not consider the extended super-Carrollian symmetry, however, we still want to investigate whether super-Carrollian symmetry has chirality. So we consider the case with two sets of supercharges as well. \par

    By solving the Jacobi identities, we find that if the algebra contains one set of generators $Q$, only $Q$'s in 2-chain representations $Q\in(j)\to(j)$ give rise to nontrivial solutions. The other cases only lead to trivial solutions. If the algebra contains two sets of generators $Q$, the solution could be nontrivial for certain pattern of two representations of $Q$'s. Especially, if both $Q_1\in(j_1), Q_2\in(j_2)$ are singlets, the solution is nontrivial for $j_1=j_2$. In the following, we discuss the results in detail. \par
    
    To be more precise, we denote supercharges as $Q_{a,\lambda}^{j,s}$, where $a=1,2$ labels the supercharge set, $\lambda$ labels the order in the chain, and $j$ and $s$ label the $\mathfrak{so}(3)$ spin and its third component respectively. For example, 
    \begin{equation}
        \begin{matrix}
            \mbox{first chain representation: }&(1/2)&\to&(3/2)&\to&(5/2)\\
            &\rotatebox{90}{$\in$} & & \rotatebox{90}{$\in$} & &\rotatebox{90}{$\in$} \\
            \mbox{supercharge examples: } &Q_{1,1}^{\frac{1}{2},\frac{1}{2}} & & Q_{1,2}^{\frac{3}{2},\frac{1}{2}} & & Q_{1,3}^{\frac{5}{2},-\frac{3}{2}} \\
        \end{matrix}
    \end{equation}
    In addition, we reorganize the bosonic generators in the raising and lowering ones:
    \begin{equation}
        \begin{aligned}
            &J^3= i J^{12}, \quad J^{\pm1}=\frac{1}{\sqrt{2}}(J^{13}\pm i J^{23}), &&B^{\pm1}=\frac{1}{\sqrt{2}}(\pm B^1+iB^2), &&P^{\pm1}=\frac{1}{\sqrt{2}}(\pm P^1+iP^2). \\
        \end{aligned}
    \end{equation}
    \par

\subsubsection{Two types of the simplest super-Carrollian algebra}

    Here we discuss two types of the simplest super-Carrollian algebra. Each one could be reproduced by taking appropriate $c\to 0$ limit from the super-Poincar\'e algebra. \par

    \paragraph{Type 1}
    The first type contains two sets of $Q_a\in(j)$ carrying the same singlet spin. The algebra contains $\{J^i, B^i, P^0, P^i, Q_{a}\}$, and the full commutation relations are
    \begin{equation}\label{eq:4dSuperCarrCase1}
        \begin{aligned}
            &[J^{s_1},J^{s_2}]=-\sqrt{2} ~\mbox{C}\smqty(1,s_1,1,s_2\\1,s_1+s_2) J^{s_1+s_2},
            \quad[J^{s_1},B^{s_2}]=-\sqrt{2} ~\mbox{C}\smqty(1,s_1;1,s_2\\1,s_1+s_2) B^{s_1+s_2},\\[0.5em]
            &[J^{s_1},P^{s_2}]=-\sqrt{2} ~\mbox{C}\smqty(1,s_1;1,s_2\\1,s_1+s_2) P^{s_1+s_2},
            \quad[B^{s_1},P^{s_2}]=\sqrt{3} ~\mbox{C}\smqty(1,s_1;1,s_2\\0,s_1+s_2) P^0,\\[0.5em]
            &[J^{s_1},Q_{a}^{j,s_2}]=\sqrt{j(j+1)} ~\mbox{C}\smqty(1,s_1;j,s_2\\j,s_1+s_2) Q_{a}^{j,s_1+s_2},
            \quad[B^{s_1},Q_{a}^{j,s_2}]=0,\\[0.5em]
            &\{Q_{1}^{j,s_1},Q_{2}^{j,s_2}\}=\mbox{C}\smqty(j,s_1;j,s_2\\0,s_1+s_2) P^0.\\
        \end{aligned}
    \end{equation}
    In the above, $\mbox{C}\smqty(j_1,s_1;j_2,s_2\\j,s)$ is the usual Clebsch-Gordan coefficients:
    \begin{equation}
        \ket{j_1,s_1}\ket{j_2,s_2}=\mbox{C}\smqty(j_1,s_1;j_2,s_2\\j,s)\ket{j,s}.
    \end{equation}
    The solution \eqref{eq:4dSuperCarrCase1} is valid for general $j\in\mathbb{Z}+1/2$. Especially, for $j=\frac{1}{2}$ the anti-commutation relations of the supercharges are
    \begin{equation}
        \{Q_1,Q_2\}=\mqty(
            0 & \frac{1}{\sqrt{2}} P^0\\
            -\frac{1}{\sqrt{2}} P^0 & 0\\), \qquad \{Q_1,Q_1\}=\{Q_2,Q_2\}=0.
    \end{equation}
    For general $j$, the anti-commutation relations can be written as 
    \begin{equation}\label{eq:4dSuperCarrQQComCase1}
        \{Q_a,Q_b\}=\mqty(
            0 & P^0 \mathbf{m}\\
            P^0 \mathbf{m}^{\mbox{\small T}}& 0\\)
            =\mqty(
            0 & P^0 \mathbf{m}\\
            - P^0 \mathbf{m}& 0\\),
    \end{equation}
    where the matrix $\mathbf{m}$ is given in term of the Clebsch-Gordan coefficients, and $\mathbf{m}^{\mbox{\small T}}=-\mathbf{m}$. \par
    
    One may wonder whether there is equivalent structure of anti-commutators by re-organizing the supercharges. The answer is no. To keep the supercharges well-behaved under rotations, the re-organization of the supercharges can be given by 
    \begin{equation}
        \tilde{Q}_a = \mathbf{M}_{ab} ~ Q_b,
    \end{equation}
    where $\mathbf{M}\in SU(2)$ is a $2$-dimensional matrix. Noticing the fact that the $\mathfrak{su}(2)$ generators, Pauli matrices $\sigma^i_{ab}$, satisfy
    \begin{equation}\label{eq:4dSuperCarrReorganizeCase1}
        \{\tilde{Q}_a, \tilde{Q}_b\}=\sigma^i_{ac} ~ \sigma^i_{bd}~\mqty(0&P^0 \mathbf{m}\\-P^0 \mathbf{m}&0)_{cd} ~ \propto ~ \mqty(0&P^0 \mathbf{m}\\-P^0 \mathbf{m}&0)_{ab},
    \end{equation}
    we see that under $SU(2)$ transformation, the anti-commutation relation is invariant up to an overall factor, which can be absorbed in rescaling the supercharges. Therefore, \eqref{eq:4dSuperCarrCase1} is the only algebraic structure for two sets of $Q_a\in(j)$. \par
    
    In fact, the relations \eqref{eq:4dSuperCarrCase1} can be obtained by taking suitable limit from $4$-dimensional super-Poincar\'e algebra. For the bosonic part of the algebra, the limit is taken by first redefining the generators 
    \begin{equation}
        J^{ij}_{\mbox{\tiny C}}=J^{ij}_{\mbox{\tiny P}}, \qquad B^i_{\mbox{\tiny C}}=c J^{0i}_{\mbox{\tiny P}}, \qquad P^0_{\mbox{\tiny C}}=c P^0_{\mbox{\tiny P}}, \qquad P^i_{\mbox{\tiny C}}= P^i_{\mbox{\tiny P}},
    \end{equation}
    where $c$ is the speed of light, and the subscripts $\mbox{C}$ and  $\mbox{P}$ label the generators in the Carrollian algebra or Poincar\'e algebra. Then by taking $c\to 0$ limit in the commutation relations of Poincar\'e algebra, we reproduce the Carrollian algebra. For the fermionic generators, we may redefine the super-Poincar\'e generators $Q^{\mbox{\tiny P}},\bar{Q}_{\mbox{\tiny P}}$ by 
    \begin{equation}\label{eq:4dSuperCarrQRescaleCase1}
        Q^{\mbox{\tiny C}}_{1}=2^{-\frac{3}{4}}c^{a}(Q^{\mbox{\tiny P}}_{2},Q^{\mbox{\tiny P}}_{1}), \qquad
        Q^{\mbox{\tiny C}}_{2}=2^{-\frac{3}{4}}c^{1-a}(-\bar{Q}_{\mbox{\tiny P}}^{\dot{1}},\bar{Q}_{\mbox{\tiny P}}^{\dot{2}})
    \end{equation}
    for constant $a$. Taking $c\to 0$ limit with Carrollian supercharge $Q=\lim_{c\to 0} Q^{\mbox{\tiny C}}$ from super-Poincar\'e algebra, we find \eqref{eq:4dSuperCarrCase1} with $j=\frac{1}{2}$. Comparing with the super-Lie algebra in the Poincar\'e case, we find that \eqref{eq:4dSuperCarrCase1} can not be treated as an extended supersymmetry, even though the two supercharges have $SU(2)$ symmetry. In fact, $Q_1$ and $Q_2$ should be interpreted as left- and right-hand part of the supersymmetry. Taking the limit with the rescaling \eqref{eq:4dSuperCarrQRescaleCase1} preserves the chirality of the algebra. \par
    
    \paragraph{Type 2}
    The second type contains one set of $Q_1$ in chain representation $(j)\to(j)$. The algebra contains the generators $\{J^i, B^i, P^0, P^i, Q_{1}\}$. The commutation relations of bosonic generators are the same with those in \eqref{eq:4dSuperCarrCase1}, and the commutation relations involving supercharges are
    \begin{equation}\label{eq:4dSuperCarrCase2}
        \begin{aligned}
            &[J^{s_1},Q_{1,\lambda}^{j,s_2}]=\sqrt{j(j+1)} ~\mbox{C}\smqty(1,s_1;j,s_2\\j,s_1+s_2) Q_{1,\lambda}^{j,s_1+s_2},
            \quad [B^{s_1},Q_{1,1}^{j,s_2}]=\mbox{C}\smqty(1,s_1;1,s_2\\j,s_1+s_2) Q_{1,2}^{j,s_1+s_2},\\[0.5em]
            &\{Q_{1,1}^{j,s_1},Q_{1,1}^{j,s_2}\}=\mbox{C}\smqty(j,s_1;j,s_2\\1,s_1+s_2) P^{s_1+s_2},
            \quad \{Q_{1,1}^{j,s_1},Q_{1,2}^{j,s_2}\}=\frac{\sqrt{3}}{2} ~ \mbox{C}\smqty(j,s_1;j,s_2\\0,s_1+s_2) P^0.\\
        \end{aligned}
    \end{equation}
    This result is also valid for generic $j$. Explicitly for $j=\frac{1}{2}$, the anti-commutation relation of the supercharges are
    \begin{equation}\label{eq:4dSuperCarrQQComCase2}
        \{Q_{1,1},Q_{1,1}\}=\mqty(
            P^{+1} & \frac{1}{\sqrt{2}} P^3 \\
            \frac{1}{\sqrt{2}} P^3 & P^{-1}\\), \qquad
        \{Q_{1,1},Q_{1,2}\}=\mqty(
            0 & \frac{1}{2}\sqrt{\frac{3}{2}} P^0\\
            -\frac{1}{2}\sqrt{\frac{3}{2}} P^0 & 0\\).
    \end{equation}
    The supercharge can be re-organized by keeping the representation under Carrollian rotations invariant
    \begin{equation}\label{eq:4dSuperCarrReorganizeCase2}
        \mqty{(j)_1 \\ \downarrow \\ (j)_2} \quad \Rightarrow \quad \mqty{(j)_1 + a(j)_2 \\ \downarrow \\ (j)_2}
    \end{equation}
    for $a$ being a constant. However, one can check that under this re-definition the anti-commutation relations are invariant, and thus \eqref{eq:4dSuperCarrQQComCase2} is the only valid algebraic structure for a single $Q_1\in(j)\to(j)$. \par

    The solution \eqref{eq:4dSuperCarrQQComCase2} can also be found by taking the ultra-relativistic limit. We should rescale the supercharges differently this time
    \begin{equation}\label{eq:4dSuperCarrQRescaleCase2}
        Q^{\mbox{\tiny C}}_{1,1}=2^{-\frac{5}{4}}c^{0}(Q^{\mbox{\tiny P}}_{2} + \bar{Q}_{\mbox{\tiny P}}^{\dot{1}},Q^{\mbox{\tiny P}}_{1} - \bar{Q}_{\mbox{\tiny P}}^{\dot{2}}), \qquad
        Q^{\mbox{\tiny C}}_{1,2}=3^{\frac{1}{2}}2^{-\frac{9}{4}}c ~ (Q^{\mbox{\tiny P}}_{2} - \bar{Q}_{\mbox{\tiny P}}^{\dot{1}},Q^{\mbox{\tiny P}}_{1} + \bar{Q}_{\mbox{\tiny P}}^{\dot{2}}).
    \end{equation}
    Taking $c\to 0$ limit and using the above rescaling lead to \eqref{eq:4dSuperCarrQQComCase2}. In this case, there is only one set of supercharge, and there is no left-right relation. Thus taking the limit breaks the chirality. \par

    Taking a limit from the super-Poincar\'e algebra can generically result in trivial solutions with $\{Q,Q\}=0$. For instance, rescaling $Q_1$ and $Q_2$ in different powers of $c$ results in a trivial super-Carrollian algebra. Besides, there are many solutions which can not be found by taking a limit. The solutions \eqref{eq:4dSuperCarrCase1} and \eqref{eq:4dSuperCarrQQComCase2} for $j\neq \frac{1}{2}$ are two examples. In the next subsection, we shall construct more solutions, all of which do not result from taking a limit. \par

\subsubsection{More complicated super-Carrollian algebras}

   There could be more complicated super-Carrollian algebras based the supercharges in longer chain representation. Here we consider two sets of supercharge $Q_1$ and $Q_2$ in chain representations respectively. After calculating over 1000 configurations, we find the following pattens leading to nontrivial solutions. It should be stressed that here we consider the singlet \eqref{eq:4dCarrSinglet} as a special case of the raising or lowering chains with length $1$. The following result applies to the singlet case as long as it satisfies the corresponding patterns. \par

    \paragraph{$Q_1$ in 2-chain} 
    Without losing generality, consider $Q_1$ in 2-chain representation. The anti-commutation relation of $Q_1$ with itself have nontrivial structure like \eqref{eq:4dSuperCarrQQComCase2}. Here we discuss other algebraic structures. \par

\begin{figure}[htpb]
        \centering
        \includegraphics[height=2cm]{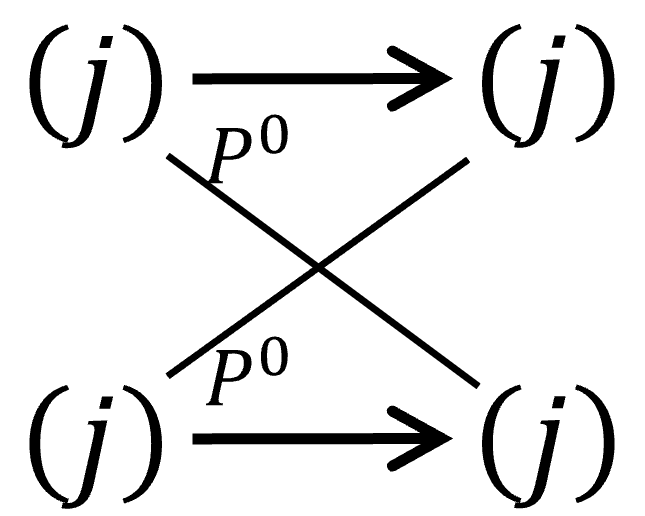}
        \caption{\centering Two 2-chain supercharges with anti-commutation relation \eqref{eq:4dSuperConCarrdouble2Chain}. The lines labels nonvanishing anti-commutators proportional to $P^0$. }
        \label{fig:4dSuperCarr2Chain2Chain}
    \end{figure}

    Let us first consider the case that $Q_2$ is in 2-chain representation as well. If both $Q_a$ are in 2-chain representation $(j_a)\to(j_a)$, there is nontrivial structure for $j_1=j_2$, and the anti-commutation relations are
    \begin{equation}\label{eq:4dSuperConCarrdouble2Chain}
        \{Q_{1,1}^{j,s_1},Q_{2,2}^{j,s_2}\}=-\{Q_{1,2}^{j,s_1},Q_{2,1}^{j,s_2}\}=a ~ \mbox{C}\smqty(j,s_1;j,s_2\\0,s_1+s_2) P^0.
    \end{equation}
    After re-organizing the $Q_a$ generators like \eqref{eq:4dSuperCarrReorganizeCase1} and \eqref{eq:4dSuperCarrReorganizeCase2}, there is one constant $a$ left undetermined, thus this super-Carrollian algebra has $1$ degree of freedom. For example, if $j=\frac{1}{2}$, we find the anti-commutation relation
    \begin{equation}
        \{Q_1,Q_2\}= \mqty(
            0 & 0 & 0 & aP^0\\
            0 & 0 & -aP^0 & 0\\
            0 & aP^0 & 0 & 0\\
            -aP^0 & 0 & 0 & 0\\).
    \end{equation}
    We can represent it as a diagram shown in Figure \ref{fig:4dSuperCarr2Chain2Chain}. \par

\begin{figure}[htpb]
        \centering
        \captionsetup[subfloat]{width=1 \linewidth} 
        \begin{minipage}[]{0.30 \linewidth}
            \subfloat[The first $\mathfrak{so}(3)$ sector of raising or lowering chain have same spin with 2-chain. \label{fig:4dSuperCarr2ChainRL:a}]{
                \qquad\includegraphics[height=1.5cm]{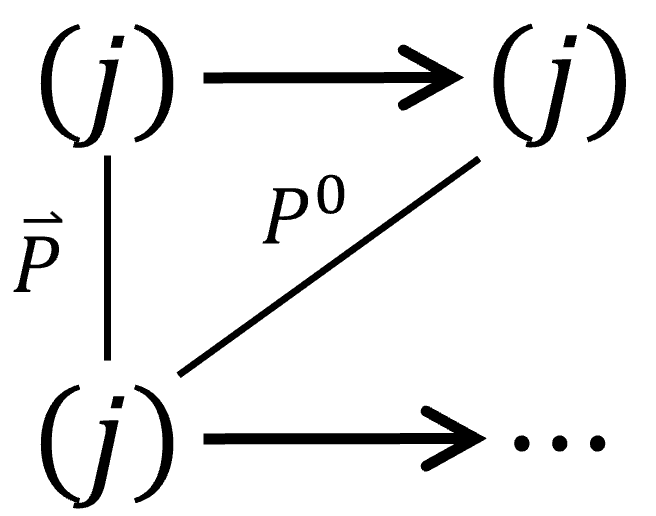}}
        \end{minipage}~
        \begin{minipage}[]{0.30 \linewidth}
            \subfloat[The second $\mathfrak{so}(3)$ sector of raising chain have same spin with 2-chain.]{
                \includegraphics[height=1.5cm]{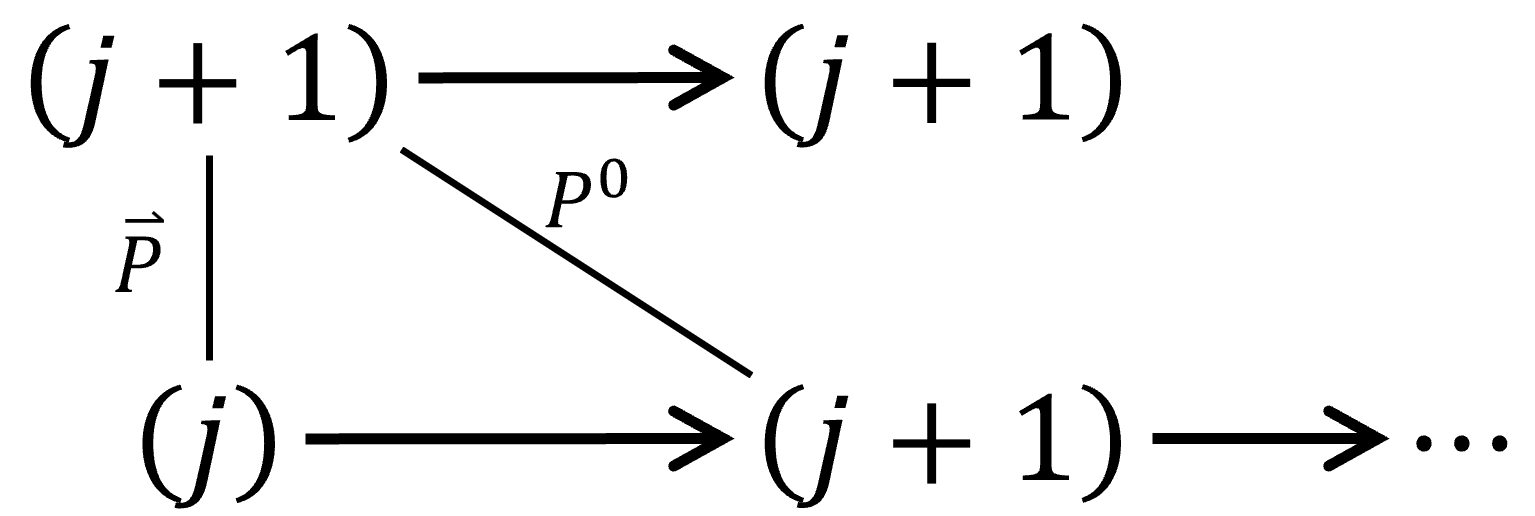}}
        \end{minipage}~
        \begin{minipage}[]{0.30 \linewidth}
        \subfloat[The second $\mathfrak{so}(3)$ sector of lowering chain have same spin with 2-chain.]{
            \includegraphics[height=1.5cm]{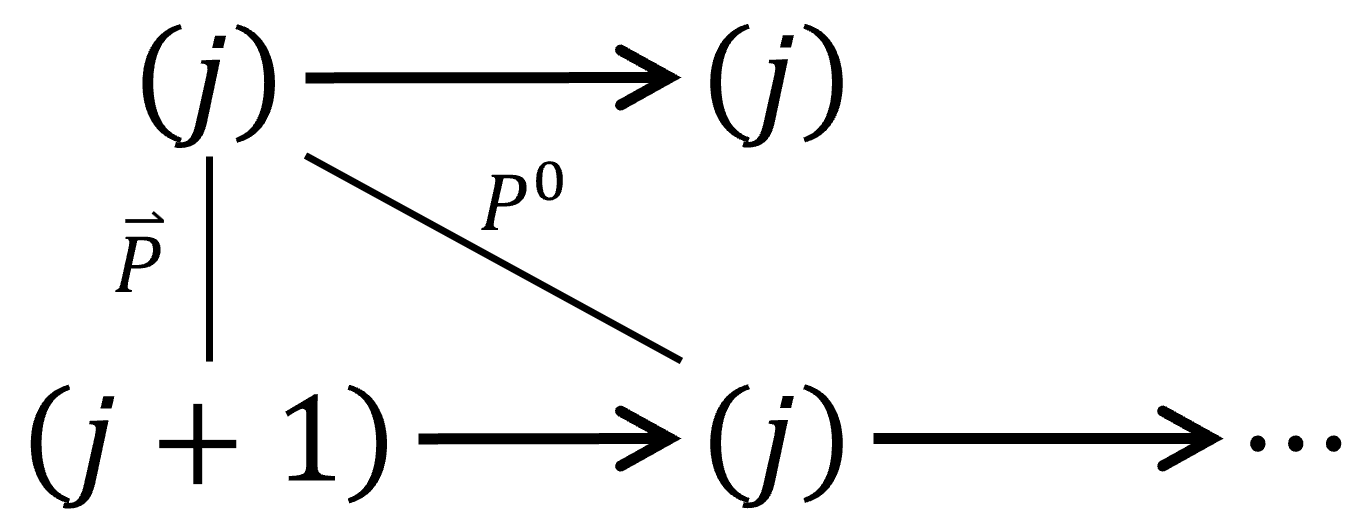}}
        \end{minipage}
        \caption{\centering One 2-chain and one raising or lowering chain supercharges. For two supercharges satisfying the pattern, the lines labels their anti-commutators proportional to $P^0$ or $P^i$ }
        \label{fig:4dSuperCarr2ChainRL}
    \end{figure}

    Next we consider the case that $Q_2$ is in raising or lowering chain representation. Now there are three kinds of structures, which are presented in Figure \ref{fig:4dSuperCarr2ChainRL}. By re-organizing the supercharges, all three structures have no degree of freedom. In particular, if $Q_2$ is in singlet representation, a special case of chain length 1, the corresponding anti-commutation relation structure is given by \ref{fig:4dSuperCarr2ChainRL:a}.\par

\begin{figure}[htpb]
        \centering
        \captionsetup[subfloat]{width=0.8 \linewidth} 
        \begin{minipage}[c][][l]{1 \linewidth}
            \subfloat[The first $\mathfrak{so}(3)$ sectors of $Q_1$ and $Q_2$ have same spin, no matter in raising or lowering chain respectively. \label{fig:4dSuperCarrRL11}]{
                \qquad\qquad\qquad\qquad\qquad\qquad\qquad\qquad
                \includegraphics[height=1.5cm]{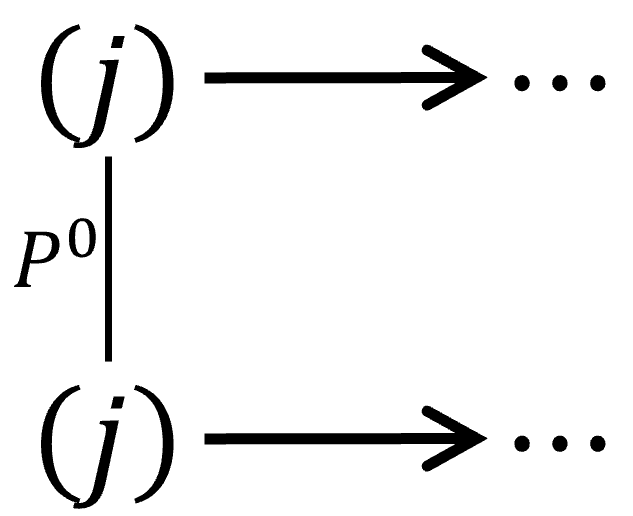}
                \qquad\qquad\qquad\qquad\qquad\qquad\qquad\qquad\qquad}
        \end{minipage}\\[1em]
        \begin{minipage}[c][][l]{1 \linewidth}
            \centering
            \subfloat[The first $n$ $\mathfrak{so}(3)$ sectors of the lowering chain is the inverse of $2$ to $n+1$ sectors of raising chain. The pattern is framed by red line. \label{fig:4dSuperCarrRLCase1}]{
                \includegraphics[height=1.5cm]{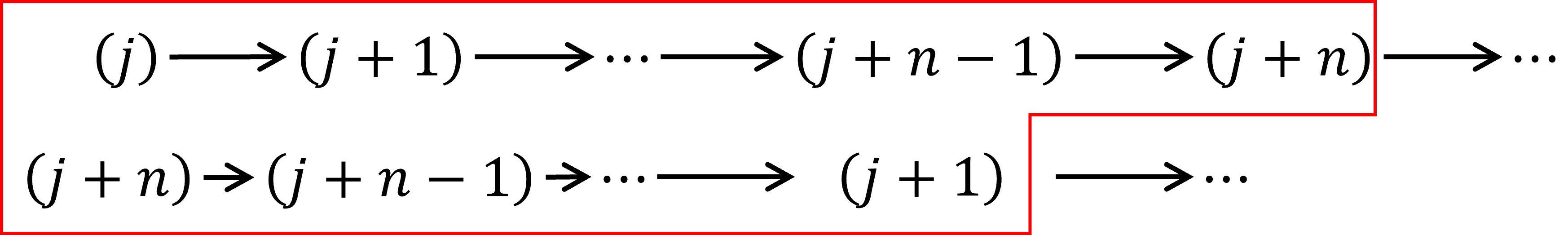}}
        \end{minipage}\\[1em]
        \begin{minipage}[c][][l]{1 \linewidth}
            \centering
            \subfloat[The first $n$ $\mathfrak{so}(3)$ sectors of the raising chain is the inverse of $2$ to $n+1$ sectors of lowering chain. The pattern is framed by blue line. \label{fig:4dSuperCarrRLCase2}]{
                \includegraphics[height=1.5cm]{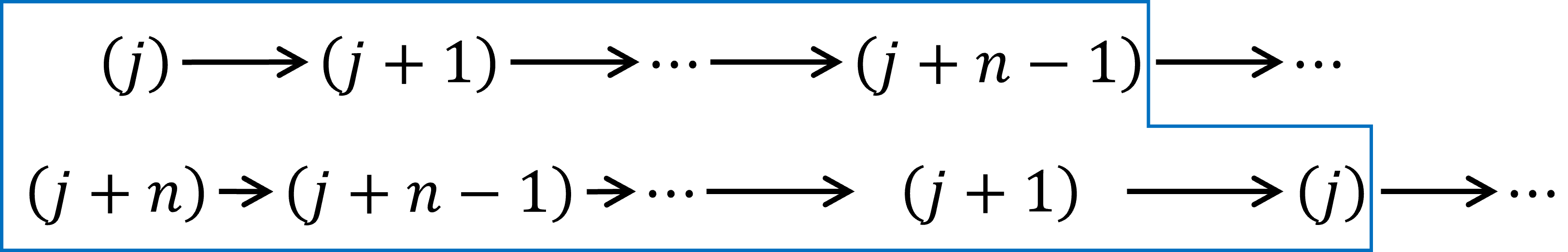}}
        \end{minipage}
        \caption{\centering Three patterns for nontrivial super-Lie algebra with $Q_1$ and $Q_2$ in raising or lowering chains. }
        \label{fig:4dSuperCarrRLCases}
    \end{figure}

\begin{figure}[htpb]
        \centering
        \includegraphics[height=1.8cm]{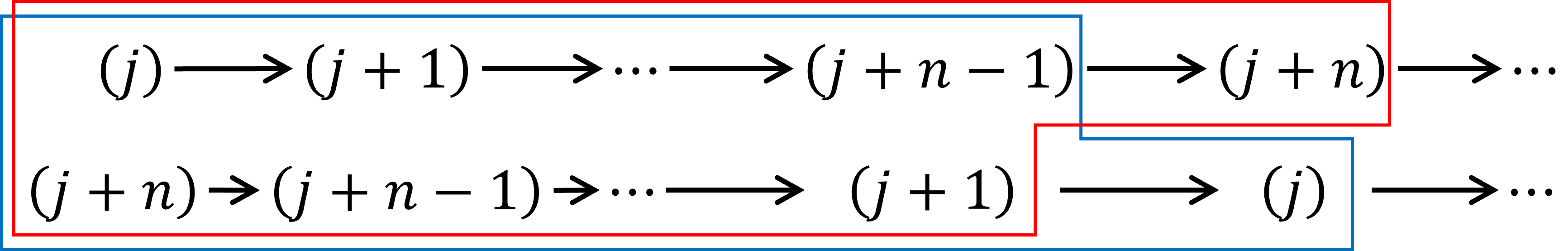}
        \caption{\centering The special case satisfying both patterns in Figure \ref{fig:4dSuperCarrRLCase1} with in red frame and \ref{fig:4dSuperCarrRLCase2} with in blue frame. Hence in this case, there is one more d.o.f. unfixed. }
        \label{fig:4dSuperCarrRLCase3}
    \end{figure}

    \paragraph{$Q_1$ and $Q_2$ in raising or lowering chains}
    If both $Q_1$ and $Q_2$ are in raising or lowering chain representations, there are three kinds of structures. If the fist sector of each chain has the same spin, no matter being raising or lowering, their anti-commutator give $P^0$, as shown in Figure \ref{fig:4dSuperCarrRL11}. For the case that $Q_1$ is in raising chain and $Q_2$ is in lowering chain, if they have the patten in Figure \ref{fig:4dSuperCarrRLCase1} within the red frame or Figure \ref{fig:4dSuperCarrRLCase2} within the blue frame, there exist nontrivial structures. Especially, if $Q_1$ and $Q_2$ have both structures of red and blue frame, the solution is nontrivial and have 1 unfixed d.o.f. as shown in Figure \ref{fig:4dSuperCarrRLCase3}. \par
    
    As an explicit example, we have 
    \begin{equation}
        \begin{aligned}
            \{Q_1,Q_2\} &= \mqty(
                \va{P} & aP^0 & 0 \\
                -(\sqrt{3}+a) P^0 & 0 & 0 \\
                0 & 0 & 0 \\), \qquad
            \mbox{with} \quad 
            \begin{aligned}
                & Q_1 \in \left(\frac{3}{2}\right)\to\left(\frac{5}{2}\right)\to\left(\frac{7}{2}\right),\\[0.5em]
                &Q_2 \in \left(\frac{5}{2}\right)\to\left(\frac{3}{2}\right)\to\left(\frac{1}{2}\right).
                \begin{tikzpicture}[overlay]
                    \draw[thick, red] (-3.9,2) -- (-1.6,2) -- (-1.6,0.7) -- (-3.0,0.7) -- (-3.0,-0.45) -- (-3.9,-0.45) -- (-3.9,2) ;
                    \draw[thick, blue, shift={(0,-0.08)}] (-3.98,2) -- (-3.0,2) -- (-3.0,0.7) -- (-1.62,0.7) -- (-1.62,-0.45) -- (-3.98,-0.45) -- (-3.98,2) ;
                \end{tikzpicture}
            \end{aligned}
        \end{aligned}
    \end{equation}
    In this expressions, we omitted the CG coefficients. The overall coefficient has been absorbed by rescaling the supercharges, and there is one undetermined coefficient $c$ which can not be fixed by re-organizing the supercharges.\par

\subsection{Carrollian Conformal supersymmetric algebra in \texorpdfstring{$d=4$}{d=4}}\label{subsec:4dSuperConCarr}

    To preserve basic structure of conformal superalgebra, we assume that apart from the Carrollian conformal algebra \eqref{eq:ConCarrAlg}, the supercharges are $Q$ and $S$ in certain representations under Carrollian rotations. Now let us assume that there is no R-symmetry and check if there is any self-contained solution. Thus the commutation relations are assumed to be
    \begin{equation}\label{eq:4dsuperConCarrAnsatz}
        \begin{aligned}
            &[D,Q]=\frac{1}{2}Q, \quad [D,S]=-\frac{1}{2}S, \quad [P,S]\propto Q, \quad [K,Q]\propto S, \\
            &\{Q_1,Q_1\}\propto P, \quad \{S_1,S_1\}\propto K, \quad\{Q,S\}=\# D+\#J+\# B.
        \end{aligned}
    \end{equation}
    \par

\subsubsection{Relation to super-Poincar\'e algebra}

    By solving various configurations, we find only three kinds of configurations\footnote{
        It should be stressed that we do not discuss extended superalgebra.
    } leading to well-defined super-Lie algebra: 
    \begin{align}
            & Q_1\in \left(\frac{1}{2}\right), \quad S_1\in \left(\frac{1}{2}\right),\label{eq:4dSuperConCarrSolution1}\\[1em]
            & Q_1\in \left(\frac{1}{2}\right)\to\left(\frac{1}{2}\right), \quad S_1\in \left(\frac{1}{2}\right)\to\left(\frac{1}{2}\right),\label{eq:4dSuperConCarrSolution2}\\[1em]
            & Q_1, Q_2\in \left(\frac{1}{2}\right), \quad S_1, S_2\in \left(\frac{1}{2}\right).\label{eq:4dSuperConCarrSolution3}
    \end{align}
    Among these, the cases \eqref{eq:4dSuperConCarrSolution1} and \eqref{eq:4dSuperConCarrSolution2} are trivial in the sense that anti-commutation relations are all vanishing $\{Q_a,Q_b\}=\{S_a,S_b\}=\{Q_a,S_b\}=0$, while the case \eqref{eq:4dSuperConCarrSolution3} is nontrivial. The commutation relations involving supercharges for \eqref{eq:4dSuperConCarrSolution3} are
    \begin{equation}\label{eq:4dSuperConCarrCommutators}
        \begin{aligned}
            &[D,Q_a]=\frac{1}{2}Q_a, \quad 
            [J^{s_1},Q_a^{\frac{1}{2},s_2}]=-\frac{\sqrt{3}}{2} ~ \mbox{C}\smqty(1,s_1;\frac{1}{2},s_2\\\frac{1}{2},s_1+s_2) Q_a^{\frac{1}{2},s_1+s_2}, \\[0.5em]
            &[D,S_a]=-\frac{1}{2}S_a, \quad 
            [J^{s_1},S_a^{\frac{1}{2},s_2}]=-\frac{\sqrt{3}}{2} ~ \mbox{C}\smqty(1,s_1;\frac{1}{2},s_2\\\frac{1}{2},s_1+s_2) S_a^{\frac{1}{2},s_1+s_2}, \\[0.5em]
            &[P^{s_1},S_a^{\frac{1}{2},s_2}] = -\sqrt{3} ~ \epsilon_{ab} \mbox{C}\smqty(1,s_1;\frac{1}{2},s_2\\\frac{1}{2},s_1+s_2) Q_a^{\frac{1}{2},s_1+s_2}, \\[0.5em]
            &[K^{s_1},Q_a^{\frac{1}{2},s_2}] = -\sqrt{3} ~ \epsilon_{ab} \mbox{C}\smqty(1,s_1;\frac{1}{2},s_2\\\frac{1}{2},s_1+s_2) S_a^{\frac{1}{2},s_1+s_2}, \\[0.5em]
            &\{Q_1^{\frac{1}{2},s_1},Q_2^{\frac{1}{2},s_2}\}= \mbox{C}\smqty(\frac{1}{2},s_1;\frac{1}{2},s_2\\0,s_1+s_2) P^0, \quad 
            \{S_1^{\frac{1}{2},s_1},S_2^{\frac{1}{2},s_2}\}= \mbox{C}\smqty(\frac{1}{2},s_1;\frac{1}{2},s_2\\0,s_1+s_2) K^0,\\[0.5em]
            &\{Q_1^{\frac{1}{2},s_1},S_2^{\frac{1}{2},s_2}\}= \mbox{C}\smqty(\frac{1}{2},s_1;\frac{1}{2},s_2\\1,s_1+s_2) B^{s_1+s_2}. \\
        \end{aligned}
    \end{equation}
    Together with the commutators of bosonic generators \eqref{eq:ConCarrAlg}, it gives the full algebra structure of $4$D Carrollian superconformal algebra. More explicitly, the anti-commutators of the supercharges are:
    \begin{equation}\label{eq:4dSuperConCarrNonTrivialSolution}
        \begin{aligned}
            &\{Q_1,Q_2\}=\mqty(0 & \frac{1}{\sqrt{2}} P^0\\-\frac{1}{\sqrt{2}} P^0 & 0\\), \quad
            &\{S_1,S_2\}=\mqty(0 & \frac{1}{\sqrt{2}} K^0\\-\frac{1}{\sqrt{2}} K^0 & 0\\), \\[1em]
            &\{Q_a,S_b\}=\delta_{ab}\mqty(B^{+1} & \frac{1}{\sqrt{2}} B^3\\\frac{1}{\sqrt{2}} B^3 & B^{-1}\\).
        \end{aligned}
    \end{equation}
    
    This result is surprising, since it shows that the Carrollian superconformal algebra does not need R-symmetry, very different from the usual superconformal algebra. On the other hand, it is reasonable because the $d$-dimensional Carrollian conformal algebra is isomorphic to $(d+1)$-dimensional Poincar\'e algebra, and its supersymmetric generalization is nothing but super-Poincar\'e algebra. Indeed, by re-organizing the bosonic generators as
    \begin{equation}
        \begin{aligned}
            &J_{\mbox{\tiny P}}^{01}=D, \quad J_{\mbox{\tiny P}}^{02}=\frac{i}{2}(K^2+P^2), \quad J_{\mbox{\tiny P}}^{03}=\frac{i}{2}(K^1+P^1), \quad J_{\mbox{\tiny P}}^{04}=-\frac{i}{2}(K^3+P^3), \\
            &J_{\mbox{\tiny P}}^{12}=\frac{i}{2}(K^2-P^2), \quad J_{\mbox{\tiny P}}^{13}=\frac{i}{2}(K^1-P^1), \quad J_{\mbox{\tiny P}}^{14}=-\frac{i}{2}(K^3-P^3), \quad\\
            &J_{\mbox{\tiny P}}^{23}=-J^{12}, \quad J_{\mbox{\tiny P}}^{24}=-J^{23}, \quad J_{\mbox{\tiny P}}^{34}=-J^{13}, \\
            &P_{\mbox{\tiny P}}^{0}=\frac{i}{\sqrt{2}}(K^0+P^0), \quad P_{\mbox{\tiny P}}^{1}=\frac{i}{\sqrt{2}}(K^0-P^0), \\
            &P_{\mbox{\tiny P}}^{2}=\sqrt{2}B^2, \quad P_{\mbox{\tiny P}}^{3}=\sqrt{2}B^1, \quad P_{\mbox{\tiny P}}^{4}=-\sqrt{2}B^3,
        \end{aligned}
    \end{equation}
    and re-organizing the fermionic generators as
    \begin{equation}
        \begin{aligned}
            &Q^{\mbox{\tiny P}}=\sqrt{2}\mqty( 
                -Q_{1,1}^{\frac{1}{2},\frac{1}{2}} + i S_{2,1}^{\frac{1}{2},-\frac{1}{2}}\\[0.5em]
                -Q_{1,1}^{\frac{1}{2},\frac{1}{2}} - i S_{2,1}^{\frac{1}{2},-\frac{1}{2}}\\[0.5em]
                -iQ_{1,1}^{\frac{1}{2},-\frac{1}{2}} + S_{2,1}^{\frac{1}{2},\frac{1}{2}}\\[0.5em]
                i Q_{1,1}^{\frac{1}{2},-\frac{1}{2}} + S_{2,1}^{\frac{1}{2},\frac{1}{2}}\\
            ), \qquad
            \bar{Q}^{\mbox{\tiny P}}=\sqrt{2}\mqty( 
                Q_{2,1}^{\frac{1}{2},\frac{1}{2}} - i S_{1,1}^{\frac{1}{2},-\frac{1}{2}}\\[0.5em]
                -Q_{2,1}^{\frac{1}{2},\frac{1}{2}} - i S_{1,1}^{\frac{1}{2},-\frac{1}{2}}\\[0.5em]
                -iQ_{2,1}^{\frac{1}{2},-\frac{1}{2}} + S_{1,1}^{\frac{1}{2},\frac{1}{2}}\\[0.5em]
                -iQ_{2,1}^{\frac{1}{2},-\frac{1}{2}} - S_{1,1}^{\frac{1}{2},\frac{1}{2}}\\
            ),
        \end{aligned}
    \end{equation}
    we reproduce the $5$D super-Poincar\'e algebra. In particular, the anti-commutator of the supercharge is
    \begin{equation}
        \begin{aligned}
            &\{Q^{\mbox{\tiny P}}, \bar{Q}^{\mbox{\tiny P}}\} = 2\vb{\Gamma}^{(1,4)}_A P_{\mbox{\tiny P}}^A,
        \end{aligned}
    \end{equation}
    where $A=0,...,4$, the subscript $P$ denotes Poincar\'e, and $\vb{\Gamma}^{(1,4)}$ are the Dirac Gamma matrices in $5$D Minkowski space, 
    \begin{equation}
        \begin{aligned}
            &\vb{\Gamma}^{(1,4)}_0=\mqty(0&\sigma^0\\\sigma^0 &0\\), \qquad
            \vb{\Gamma}^{(1,4)}_i=\mqty(0&-\sigma^i\\\sigma^i &0\\), \\[1em]
            &\vb{\Gamma}^{(1,4)}_4=\vb{\Gamma}^{(1,4)}_0\vb{\Gamma}^{(1,4)}_1\vb{\Gamma}^{(1,4)}_2\vb{\Gamma}^{(1,4)}_3 
            = \mqty(-i\sigma^0 &0\\0 &i\sigma^0 \\),
        \end{aligned}
    \end{equation}
    and our convention of Pauli matrices are given by 
    \begin{equation}
        \sigma^0=\mqty(1&0\\0&1\\), \quad \sigma^1=\mqty(0&1\\1&0\\), \quad \sigma^2=\mqty(0&-i\\i&0\\), \quad \sigma^3=\mqty(1&0\\0&-1\\).
    \end{equation}
    \par

    Form the structure of \eqref{eq:4dSuperConCarrSolution3}, one may suspect that there exist $SU(2)$ symmetry between $Q_1$ and $Q_2$, as well as between $S_1$ and $S_2$. This is true. However, as will be explained in Section \ref{subsubsec:4dSuperConCarrRsymmetry} shortly, this $SU(2)$ symmetry is just an outer automorphism, and does not appear in $\{Q_a,S_b\}$ anti-commutation relations. More precisely, the algebra here is similar to the case of \eqref{eq:4dSuperCarrCase1}, where the $Q_1$ and $Q_2$ should be viewed as left- and right-hand sectors of the superalgebra, and the $SU(2)$ is purely a redefinition of left- and right-hand sectors. Denoting the generators as $T^i$ with $i = 1, 2, 3$, the supercharges transform under $SU(2)$ as
    \begin{equation}\label{eq:4dSuperConCarrSU2}
        [T^i,Q_a]=\sigma^i_{ab}Q_b, \qquad [T^i,S_a]=\sigma^i_{ab}S_b.
    \end{equation}
    From the point of view of the $5$D super-Poincar\'e algebra , the representations are organized using Cartan subalgebra $\{J^{01}_{\mbox{\tiny P}}, J^{23}_{\mbox{\tiny P}}\}$. The $SU(2)=SO(3)$ transformation is equivalent to rotating $2, 3, 4$ directions, exchanging $J^{23}_{\mbox{\tiny P}}$ to another rotation in the Cartan subalgebra, as well as interchange $Q^{\mbox{\tiny P}}$ and $\bar{Q}^{\mbox{\tiny P}}$. Under this symmetry, the super-Lie algebra is equivalent up to the corresponding change of Cartan subalgebra. \par
    
    To conclude, the nontrivial Carrollian superconformal algebra in $4$D is isomorphic to super-Poincar\'e algebra in $5$D. \par

\subsubsection{Comments on R-symmetry}\label{subsubsec:4dSuperConCarrRsymmetry}

    As mentioned in Section \ref{subsec:SuperConformal}, the superconformal algebra is closed only by including the R-symmetry. Introducing R-symmetry in the discussion of Carrollian case, we find no fundamental change to the structure of the algebra of \eqref{eq:4dSuperConCarrSolution1}, \eqref{eq:4dSuperConCarrSolution2} and \eqref{eq:4dSuperConCarrSolution3}. More explicitly, we change the ansatz in such a way that $Q_a$ and $S_a$ carry some R charges, and their anti-commutators contains R-symmetry generators $R$
    \begin{equation}
        \begin{aligned}
            &[R,Q_a]=r_{Q_a}Q_a, \qquad [R,S_a]=r_{S_a}S_a, \\
            &\{Q_1,S_1\}=\# D+\#J+\# B + \# R.
        \end{aligned}
    \end{equation}
    
    In the case \eqref{eq:4dSuperConCarrSolution1}, $R$ generates $U_R(1)$ symmetry. In a well-defined super-Lie algebra, $Q_1$ and $S_1$ carry the same R charges $r_{Q_1}=r_{S_1}=r$, while their anti-commutation relations are vanishing
    \begin{equation}
        [R,Q_1]=rQ_1, \qquad [R,S_1]=rS_1, \qquad \{Q_1,Q_1\}=\{S_1,S_1\}=\{Q_1,S_1\}=0.
    \end{equation}
    Hence, this structure is trivial. \par

    In the case of \eqref{eq:4dSuperConCarrSolution2}, $R$ generates $U_R(1)$ as well. There are two kinds of structures. The first one is trivial, similar to the case of \eqref{eq:4dSuperConCarrSolution1}: the R charges are the same $r_{Q_1}=r_{S_1}=r$, and $R$ does not appear in the anti-commutators of the supercharges
    \begin{equation}
        [R,Q_a]=r Q_a, \quad [R,S_a]=r S_a, \quad \{Q_a,Q_b\}=\{S_a,S_b\}=\{Q_a,S_b\}=0.
    \end{equation}
    The other one is that R charges are vanishing $r_{Q_1}=r_{S_1}=0$, while the anti-commutation relations are nonvanishing:
    \begin{equation}
        \{Q_1,Q_1\}=\{S_1,S_1\}=0, \qquad \{Q_1,S_1\}\supset R.
    \end{equation}
    In this, the R-symmetry commutes with all other generators, and can be viewed as a central charge. \par

    In the case of \eqref{eq:4dSuperConCarrSolution3}, R-symmetry is $U_R(1)\otimes SU_R(2)$, and we denote the corresponding generators as $R$ and $T^i$, respectively. Since $[R,T^i]=0$, we can discuss them separately. For the $U_R(1)$ part, the situation is similar to the case of \eqref{eq:4dSuperConCarrSolution2}: either $R$ is a trivial $U(1)$ symmetry with 
    \begin{equation}\label{eq:4dSuperConCarrRTrivial}
        \begin{aligned}
            &[R,Q_a]=r \epsilon_{ab}Q_b, \quad [R,S_a]=-r \epsilon_{ab}S_b, \\
            &\{Q_a,Q_b\}\not\supset R \quad \{S_a,S_b\}\not\supset R, \quad\{Q_a,S_b\}\not\supset R.
        \end{aligned}
    \end{equation}
    and the anti-commutation relations of supercharges are the same as \eqref{eq:4dSuperConCarrNonTrivialSolution}; or $R$ is a central charge with 
    \begin{equation}\label{eq:4dSuperConCarrRCentralCharge}
        \{Q_a,S_b\}\supset R, \quad [R,G]=0, ~\forall G.
    \end{equation}
    For the $SU_R(2)$ part, it is quite reasonable to assume \eqref{eq:4dSuperConCarrSU2} being valid. Solving the Jacobi identities with the ansatz $\{Q_a,S_b\}=\# D+\#J+\# B + \# T$, we find that $T^i$'s do not appear in the anti-commutators. Thus this $SU(2)$ symmetry is just an outer automorphism. Moreover, $T^i$'s do not appear in $\{Q_a,S_b\}$ even assuming $[T^i,Q_a]=[T^i,S_a]=0$. \par

    We also investigated the cases that the supercharges are in other chain representations with different spin $j$, but found no more structure even after introducing R-symmetry. So the structure of Carrollian superconformal algebra in $4$D is determined uniquely and can not be modified by introducing R-symmetry. \par

\subsubsection{Nontrivial algebra from taking ultra-relativistic limit}

    The algebra \eqref{eq:4dSuperConCarrCommutators} can also be found by taking the ultra-relativistic limit from superconformal algebra with a rescaling similar to \eqref{eq:4dSuperCarrQRescaleCase1}. To be precise, the rescaling of the supercharges are
    \begin{equation}
        \begin{aligned}
            &Q^{\mbox{\tiny C}}_{1}=2^{-\frac{3}{4}}c^{a}(Q^{\mbox{\tiny P}}_{2},Q^{\mbox{\tiny P}}_{1}), 
            &&Q^{\mbox{\tiny C}}_{2}=2^{-\frac{3}{4}}c^{1-a}(-\bar{Q}_{\mbox{\tiny P}}^{\dot{1}},\bar{Q}_{\mbox{\tiny P}}^{\dot{2}}),\\
            &S^{\mbox{\tiny C}}_{1}=-2^{-\frac{3}{4}}c^{a}(S^{\mbox{\tiny P}}_{2},S^{\mbox{\tiny P}}_{1}), 
            &&S^{\mbox{\tiny C}}_{2}=-2^{-\frac{3}{4}}c^{1-a}(-\bar{S}_{\mbox{\tiny P}}^{\dot{1}},\bar{S}_{\mbox{\tiny P}}^{\dot{2}}),\\
        \end{aligned}
    \end{equation}
    with constant $a$. Taking $c\to 0$ limit with $Q=\lim_{c\to 0} Q^{\mbox{\tiny C}}$, we see that the superconformal algebra then produce \eqref{eq:4dSuperConCarrCommutators}. Especially, the $U_R(1)$ R-symmetry is rescaled as 
    \begin{equation}
        R^{\mbox{\tiny P}}=c^b R^{\mbox{\tiny P}}.
    \end{equation}
    To find a non-divergent result, the constant $b$ satisfy $0\le b\le 1$. There are three scenarios: for $b=0$, the R-symmetry becomes a trivial $U(1)$ symmetry as in \eqref{eq:4dSuperConCarrRTrivial}; for $b=1$, the R-symmetry becomes a central charge shown in $\{Q, S\}$ with $[R,G]=0$ for any generator $G$ in the algebra; for the case that $0<b<1$, the R-symmetry totally decouple in the sense that $R$ commutes with any other generator $[R,G]=0$, as well as not appearing in any commutator. This explains the two kinds of structures of R-symmetry found in the last subsection. \par

\section{Carrollian superconformal symmetry in \texorpdfstring{$d=3$}{d=3}}\label{sec:3dCarrSUSY}

    In this section, we discuss the $3$-dimensional case. As introduced in Section \ref{sec:CarrAlg}, the representation of $3$-dimensional Carrollian rotations is less restricted. We will not dive into the nonconformal case since we do not have preferred ansatz on the representations. On the contrary, our focus is turned to supersymmetric extension of $3$D Carrollian conformal algebra, and its relation to BMS algebra. \par

    In Section \ref{subsec:3dSuperConCarr} we shall only show the structure of $3$D Carrollian superconformal algebra which is isomorphic to $4$D super-Poincar\'e algebra. Besides, there is no R-symmetry needed in the algebra. In Section \ref{subsec:SingletSuperBMS4}, we further extend this algebra to singlet super BMS$_4$ algebra. In Section \ref{subsec:MultipletletSuperBMS4}, we demonstrate multiplet super-BMS$_4$ algebra which can not be obtained by extending Carrollian superconformal algebra. In each case, we further discuss the Hermitian conjugation conditions. 

\subsection{Carrollian superconformal algebra in \texorpdfstring{$d=3$}{d=3}}\label{subsec:3dSuperConCarr}

    As mentioned in Section \ref{sec:CarrAlg}, the $3$-dimensional Carrollian conformal algebra is isomorphic to $4$D Poincar\'e algebra $\mathfrak{cca}_3 = \mathfrak{iso}(1,3)$. By the Haag--Łopuszański--Sohnius theorem, the super-Lie algebra with nonvanishing $\{Q, Q\}$ anti-commutators is given by $Q_{\mbox{\tiny P}}$ and $\bar{Q}_{\mbox{\tiny P}}$, with commutation relation given by \eqref{eq:4dSuperPoincareAlgebra}. The $3$-dimensional Carrollian superconformal algebra has generators $\{D, J^{12}, B^i, P^\mu, K^\mu, Q_s, S_s\}$, with $\mu=0,1,2$, $i=1,2$ and spin $s=\pm\frac{1}{2}$, which can be related to the generators of the $4$-dimensional super Poincar\'e algebra by the following redefinitions:
    \begin{equation}
        \begin{aligned}
            & D=J^{03}_{\mbox{\tiny P}}, \quad J^{12}=J^{12}_{\mbox{\tiny P}}, \quad B^{1}=-i\sqrt{2} P^{2}_{\mbox{\tiny P}}, \quad B^{2}=i\sqrt{2} P^{1}_{\mbox{\tiny P}}, \\[0.5em]
            & P^0=\sqrt{2}(P^{0}_{\mbox{\tiny P}}-P^{3}_{\mbox{\tiny P}}), P^1=i(J^{02}_{\mbox{\tiny P}}+J^{23}_{\mbox{\tiny P}}), \quad P^2=-i(J^{01}_{\mbox{\tiny P}}+J^{13}_{\mbox{\tiny P}}), \\[0.5em]
            & K^0=\sqrt{2}(P^{0}_{\mbox{\tiny P}}+P^{3}_{\mbox{\tiny P}}), K^1=i(J^{02}_{\mbox{\tiny P}}-J^{23}_{\mbox{\tiny P}}), \quad K^2=i(-J^{01}_{\mbox{\tiny P}}+J^{13}_{\mbox{\tiny P}}), \\[0.5em]
            & Q_{\frac{1}{2}}=\frac{1}{\sqrt{2}}Q^{\mbox{\tiny P}}_{2}, \quad Q_{-\frac{1}{2}}=\frac{1}{\sqrt{2}} \bar{Q}^{\dot{2}}_{\mbox{\tiny P}}, \quad S_{\frac{1}{2}}=-\frac{1}{\sqrt{2}} \bar{Q}^{\dot{1}}_{\mbox{\tiny P}}, \quad S_{-\frac{1}{2}}=-\frac{1}{\sqrt{2}}Q^{\mbox{\tiny P}}_{1}.
        \end{aligned}
    \end{equation}
    Especially, the commutators of the bosonic generators are the same with \eqref{eq:ConCarrAlg}, while the commutators involving supercharges are
    \begin{equation}\label{eq:3dSuperConCarrAlg}
        \begin{aligned}
            &[D,Q_s] = \frac{1}{2} Q_s, \quad [D,S_s] = \frac{1}{2} S_s, \quad [J^{12},Q_s] = -is Q_s, \quad [J^{12},S_s] = -is S_s, \\[0.5em]
            &[P^1,S_s]=\mqty(-Q_{-\frac{1}{2}}\\Q_{\frac{1}{2}}), \quad [P^2,S_s]=\mqty(-iQ_{-\frac{1}{2}}\\-iQ_{\frac{1}{2}}), \\[0.5em]
            &[K^1,Q_s]=\mqty(-S_{-\frac{1}{2}}\\S_{\frac{1}{2}}), \quad [K^2,Q_s]=\mqty(-iS_{-\frac{1}{2}}\\-iS_{\frac{1}{2}}),\\[0.5em]
            &\{Q_{s_1},Q_{s_2}\}=\frac{1}{\sqrt{2}}\mqty(0&P^0\\P^0&0), \quad \{S_{s_1},S_{s_2}\}=\frac{1}{\sqrt{2}}\mqty(0&K^0\\K^0&0), \\[1em]
            &\{Q_{s_1},S_{s_2}\}=\frac{1}{\sqrt{2}}\mqty(B^1+iB^2 & 0\\ 0 & -B^1+iB^2).
        \end{aligned}
    \end{equation}
    It is interesting to point out that this algebra can also be found by taking $c\to0$ limit from $3$D superconformal algebra. \par

    Moreover, let us check if the introduction of the R-symmetry may make a difference. For this algebra, the supercharge has just one copy, and the R-symmetry is $U_R(1)$. Assuming 
    \begin{equation}
        [R,Q_s]\propto Q_s, \quad [R,S_s]\propto S_s, \quad \{Q_{s_1},Q_{s_2}\}\supset R, \quad \{S_{s_1},S_{s_2}\}\supset R, \quad \{Q_{s_1},S_{s_2}\}\supset R,
    \end{equation}
    we can solve the Jacobi identities. The only solution is that $R$ not appearing in anti-commutation relations 
    \begin{equation}
        \{Q_{s_1},Q_{s_2}\}\not\supset R, \quad \{S_{s_1},S_{s_2}\}\not\supset R, \quad \{Q_{s_1},S_{s_2}\}\not\supset R,
    \end{equation}
    and the fermionic generators have R charges:
    \begin{equation}\label{eq:3dConCarrRsymmetry}
        [R,Q_{\frac{1}{2}}] = rQ_{\frac{1}{2}}, \quad [R,Q_{-\frac{1}{2}}] = -rQ_{-\frac{1}{2}}, \quad [R,S_{\frac{1}{2}}] = -rS_{\frac{1}{2}}, \quad [R,S_{-\frac{1}{2}}] = rS_{-\frac{1}{2}}.
    \end{equation}
    It is worth mentioning that there is no $SU(2)$ outer automorphism in 3D nontrivial Carrollian superconformal algebra \eqref{eq:3dSuperConCarrAlg}, which is different from the $4$D case discussed in Section \ref{subsubsec:4dSuperConCarrRsymmetry}. An outer automorphism of fermionic part of the algebra is nothing but a map between equivalent representations of the Carrollian conformal algebra. Contrary to the case in Section \ref{subsubsec:4dSuperConCarrRsymmetry}, $J^{12}_{\mbox{\tiny P}}$ is the only Lorentzian rotation commuting with $J^{03}_{\mbox{\tiny P}}$. Therefore the requirement that $D=J^{03}_{\mbox{\tiny P}}$ being diagonal in the representation leads to the unique choice of Cartan subalgebra to be $\{D=J^{03}_{\mbox{\tiny P}}, J^{12}=J^{12}_{\mbox{\tiny P}}\}$. Thus the outer automorphism in 3D case can only be a $U(1)$ symmetry, which is precisely the $R$ generator in \eqref{eq:3dConCarrRsymmetry}. \par

\subsection{Singlet super-BMS\texorpdfstring{$_4$}{4} algebra}\label{subsec:SingletSuperBMS4}

    One thing makes Carrollian conformal algebra in $3$D special is that it admit infinite extensions to BMS$_4$ algebra. In fact, the above super-Lie algebra also admits infinite extension. The BMS$_4$ generators are identified with Carrollian conformal generators as
    \begin{equation}
        \begin{aligned}
            &L_1 = \frac{1}{2}(K^1-iK^2), \quad L_0 = \frac{1}{2}(D+iJ^{12}), \quad L_{-1} = \frac{1}{2}(P^1+iP^2), \\[0.5em]
            &\bar{L}_1 = \frac{1}{2}(-K^1-iK^2), \quad \bar{L}_0 = \frac{1}{2}(D-iJ^{12}), \quad \bar{L}_{-1} = \frac{1}{2}(-P^1+iP^2), \\[0.5em]
            &M_{\frac{1}{2},\frac{1}{2}} = \frac{1}{\sqrt{2}}K^0, \quad M_{\frac{1}{2},-\frac{1}{2}} = \frac{1}{\sqrt{2}}(B^1-iB^2), \\[0.5em]
            &M_{-\frac{1}{2},\frac{1}{2}} = \frac{1}{\sqrt{2}}(-B^1-iB^2), \quad M_{-\frac{1}{2},-\frac{1}{2}} = \frac{1}{\sqrt{2}}P^0. \\
        \end{aligned}
    \end{equation}
    Then the BMS$_4$ algebra is given by 
    \begin{equation}\label{eq:BMS4Algebra}
        \begin{aligned}
            &[L_m,L_n]=(m-n)L_{m+n}, \quad [\bar{L}_{\bar{m}},\bar{L}_{\bar{n}}]=(m-n)\bar{L}_{\bar{m}+\bar{n}}, \\
            &[L_m, M_{r,\bar{r}}] = \left(\frac{m}{2}-r\right)M_{m+r,\bar{r}}, \quad [\bar{L}_{\bar{m}}, M_{r,\bar{r}}] = \left(\frac{\bar{m}}{2}-\bar{r}\right)M_{r,\bar{m}+\bar{r}},
        \end{aligned}
    \end{equation}
    and is extended to $m,\bar{m}\in\mathbb{Z}, r, \bar{r}\in\mathbb{Z}+\frac{1}{2}$. The fermionic generators in super-extended BMS$_4$ algebra are denoted by curly $\mathcal{Q}$, and their relations to Carrollian superconformal generators are
    \begin{equation}
        \mathcal{Q}_{\frac{1}{2}} = -\sqrt{2} S_{-\frac{1}{2}}, \quad \mathcal{Q}_{-\frac{1}{2}} = \sqrt{2} Q_{\frac{1}{2}}, \quad \bar{\mathcal{Q}}_{\frac{1}{2}} = -\sqrt{2} S_{\frac{1}{2}}, \quad \bar{\mathcal{Q}}_{-\frac{1}{2}} = \sqrt{2} Q_{-\frac{1}{2}},
    \end{equation}
    and the generators can be extended to $\mathcal{Q}_r$ and $\bar{\mathcal{Q}}_{\bar{r}}$ with $r, \bar{r}\in\mathbb{Z}+\frac{1}{2}$. The commutators involving the fermionic generators are
    \begin{equation}\label{eq:SingletSuperBMS4}
        [L_m,\mathcal{Q}_r] = \left(\frac{m}{2}-r\right)\mathcal{Q}_{m+r}, \quad 
        [\bar{L}_{\bar{m}}, \bar{\mathcal{Q}}_{\bar{r}}] = \left(\frac{\bar{m}}{2}-\bar{r}\right)\bar{\mathcal{Q}}_{\bar{m}+\bar{r}}, \quad 
        \{\mathcal{Q}_{r}, \bar{\mathcal{Q}}_{\bar{r}}\} = M_{r,\bar{r}}.
    \end{equation}
    Together, \eqref{eq:BMS4Algebra} and \eqref{eq:SingletSuperBMS4} gives the full super-BMS$_4$ algebraic structures. For later convenience, we call this superalgebra the singlet super-BMS$_4$ algebra in the sense that $\mathcal{Q}$ and $\bar{\mathcal{Q}}$ are all singlets under $M$:
    \begin{equation}
        [M, \mathcal{Q}]=[M,\bar{\mathcal{Q}}]=0.
    \end{equation}
    \par

    It is interesting to discuss various definitions of the Hermitian conjugation conditions. Let us first restrict our attention to conjugation condition on one branch of the Virasoro algebra $L_n$. The conjugation operation $\dagger$ is defined to satisfy 
    \begin{equation}
        (aG)^\dagger = a^* G^\dagger, \qquad (G^\dagger)^\dagger = G,
    \end{equation}
    for any algebraic generator $G$ and complex constant $a$. Assuming that $L_m^\dagger = a(m) L_{f(m)}$, the above ansatz translates to 
    \begin{equation}
        a(m)^*a(f(m))=1, \qquad f(f(m))=m.
    \end{equation}
    The compatible condition with the commutation relations further requires
    \begin{equation}
        -a(m)a(n)(f(m)-f(n)) = (m-n)~a(m+n), \qquad f(m)+f(n) = f(m+n).
    \end{equation}
    There are two solutions to the above equations:
    \begin{equation}
        \begin{aligned}
            &\mbox{case 1}: && L_m^\dagger = -e^{i c m} L_{m},\\
            &\mbox{case 2}: && L_m^\dagger = e^{c m} L_{-m},\\
        \end{aligned}
    \end{equation}
    for real constant $c\in\mathbb{R}$. Thus considering both the holomorphic and anti-holomorphic sectors, we have four kinds of definition on the conjugation, in which $L/\bar{L}$ conjugates to $L/\bar{L}$. We call these type I conjugation conditions. On the other hand, we can define type II conjugation conditions in which $L$ conjugates to $\bar{L}$ and vice versa. To satisfy $(G^\dagger)^\dagger = G$, there are two kinds of type II conjugation conditions. In each case, we can fix the conjugation conditions on $M$ and supercharges by requiring compatibility with the commutators. In Table \ref{tab:SingletSuperBMS4ConjugationConditions}, we conclude all possible conjugation conditions.

    \begin{table}[htpb]
        \renewcommand\arraystretch{1.8}
        \centering
        \caption{\centering The Hermitian conjugation conditions of singlet super BMS$_4$ algebra. }
        \begin{tabular}{c|c|c|c|c|c}
            \hline
             & $L_m$ & $\bar{L}_{\bar{m}}$ & $M_{r,\bar{r}}$ & $\mathcal{Q}_{r}$ & $\bar{\mathcal{Q}}_{\bar{r}}$\\\hline\hline
            type I-11 & $- e^{i c_1 m} L_m$ & $- e^{i c_2 \bar{m}} \bar{L}_{\bar{m}}$ 
                & $e^{i c_1 r +i c_2 \bar{r}+ ic_3 + ic_4} M_{r,\bar{r}}$ 
                & $e^{i c_1 r + ic_3}\mathcal{Q}_{r}$ & $e^{i c_2 \bar{r} + ic_4}\bar{\mathcal{Q}}_{\bar{r}}$\\\hline
            type I-12 & $- e^{i c_1 m} L_m$ & $e^{c_2 \bar{m}} \bar{L}_{-\bar{m}}$ 
                & $e^{i c_1 r + c_2 \bar{r}+ ic_3 + ic_4} M_{r,-\bar{r}}$ 
                & $e^{i c_1 r+ ic_3}\mathcal{Q}_{r}$ & $e^{c_2 \bar{r} + ic_4}\bar{\mathcal{Q}}_{-\bar{r}}$\\\hline
            type I-21 & $e^{c_1 m}L_{-m}$ & $- e^{i c_2 \bar{m}} \bar{L}_{\bar{m}}$ 
                & $e^{ c_1 r +i c_2 \bar{r}+ ic_3 + ic_4} M_{-r,\bar{r}}$ 
                & $e^{c_1 r+ ic_3}\mathcal{Q}_{-r}$ & $e^{i c_2 \bar{r} + ic_4}\bar{\mathcal{Q}}_{\bar{r}}$\\\hline
            type I-22 & $e^{c_1 m}L_{-m}$ & $e^{c_2 \bar{m}} \bar{L}_{-\bar{m}}$ 
                & $e^{ c_1 r + c_2 \bar{r}+ ic_3 + ic_4} M_{-r,-\bar{r}}$ 
                & $e^{c_1 r+ ic_3}\mathcal{Q}_{-r}$ & $e^{c_2 \bar{r} + ic_4}\bar{\mathcal{Q}}_{-\bar{r}}$\\\hline\hline
            type II-1 & $- e^{i c_1 m} \bar{L}_{m}$ & $- e^{i c_1 \bar{m}} L_{\bar{m}}$ 
                & $e^{ic_1 (r+\bar{r})+2ic_2} M_{\bar{r},r}$ 
                & $e^{i c_1 r + ic_2}\bar{\mathcal{Q}}_{r}$ & $e^{i c_1 \bar{r} + ic_2}\mathcal{Q}_{\bar{r}}$\\\hline
            type II-2 & $e^{c_1 m} \bar{L}_{-m}$ & $e^{c_1 \bar{m}}L_{-\bar{m}}$ 
                & $e^{c_1 (r+\bar{r})+2ic_2} M_{-\bar{r},-r}$ 
                & $e^{c_1 r + ic_2}\bar{\mathcal{Q}}_{-r}$ & $e^{c_1 \bar{r}+ ic_2}\mathcal{Q}_{-\bar{r}}$\\\hline
        \end{tabular}
        \label{tab:SingletSuperBMS4ConjugationConditions}
    \end{table}

    Now let us identify the Hermitian conjugations in Table \ref{tab:SingletSuperBMS4ConjugationConditions} with various conjugations in the literature. 
    \paragraph{BMS$_4$ theory} As the global part of the BMS algebra gives Poincar\'e algebra, the physical conjugation condition is given by
    \begin{equation}
        (J_{\mbox{\tiny P}}^{\mu\nu})^\dagger = -J_{\mbox{\tiny P}}^{\mu\nu}, \quad (P_{\mbox{\tiny P}}^{\mu})^\dagger = P_{\mbox{\tiny P}}^{\mu}, \quad (Q^{\mbox{\tiny P}}_{\alpha})^\dagger = \bar{Q}_{\mbox{\tiny P}}^{\alpha}, \quad (\bar{Q}_{\mbox{\tiny P}}^{\dot{\alpha}})^\dagger = Q^{\mbox{\tiny P}}_{\dot{\alpha}}.
    \end{equation}
    This corresponds to type II-1 case with $c_i=0$:
    \begin{equation}
        L_{m}^\dagger = -\bar{L}_{m}, \quad \bar{L}_{\bar{m}}^\dagger = -L_{\bar{m}}, \quad M_{r,\bar{r}}^\dagger = M_{\bar{r},r}, \quad \mathcal{Q}_{r}=\bar{\mathcal{Q}}_{r}, \quad \bar{\mathcal{Q}}_{\bar{r}} = \mathcal{Q}_{\bar{r}}.
    \end{equation}
    The symmetry of celestial CFT corresponding to the $4$D Poincar\'e theory is generated by the Virasoro part of BMS$_4$ algebra. The above conjugation relation is equivalent to BPZ conjugation condition of Euclidean celestial CFT with $L_{m}^\dagger = -\bar{L}_{m}$ and $\bar{L}_{\bar{m}}^\dagger = -L_{\bar{m}}$.

    \paragraph{$2$D Lorentzian celestial CFT} For Lorentzian celestial CFT, the BPZ conjugation is given by $L_{m}^\dagger = -L_{m}$ and $\bar{L}_{\bar{m}}^\dagger = -\bar{L}_{\bar{m}}$. It corresponds to type I-11 with $c_i=0$:
    \begin{equation}
        L_{m}^\dagger = -L_m, \quad \bar{L}_{\bar{m}}^\dagger = -\bar{L}_{\bar{m}}, \quad M_{r,\bar{r}}^\dagger = M_{r,\bar{r}}, \quad \mathcal{Q}_{r}=\mathcal{Q}_{r}, \quad \bar{\mathcal{Q}}_{\bar{r}} = \bar{\mathcal{Q}}_{\bar{r}}.
    \end{equation}
    
    \paragraph{$3$D Carrollian CFT}Viewing the global part of this super-BMS$_4$ algebra as the $3$D Carrollian conformal symmetry, the BPZ conjugation condition should be
    \begin{equation}
        \begin{aligned}
            &D^\dagger=D, \quad (J^{12})^\dagger = -J^{12}, \quad (B^{i})^\dagger = -B^{i}, \quad (P^{\mu})^\dagger = K^{\mu}, \quad (K^{\mu})^\dagger = P^{\mu}, \\
            &Q_s^\dagger = -S_{-s}, \quad S_s^\dagger = -Q_{-s}.
        \end{aligned}
    \end{equation}
    The corresponding conjugation condition on super-BMS$_4$ algebra is type I-11 with $c_1=c_2=\pi, c_3=c_4=\frac{\pi}{2}$:
    \begin{equation}
         \begin{aligned}
             &L_{m}^\dagger = (-1)^{m+1}L_m, \quad \bar{L}_{\bar{m}}^\dagger = (-1)^{\bar{m}+1}\bar{L}_{\bar{m}}, \quad M_{r,\bar{r}}^\dagger = (-1)^{r+\bar{r}+1}M_{r,\bar{r}}, \\
             &\mathcal{Q}_{r}=(-1)^{r+\frac{1}{2}}\mathcal{Q}_{r}, \quad \bar{\mathcal{Q}}_{\bar{r}} = (-1)^{\bar{r}+\frac{1}{2}}\bar{\mathcal{Q}}_{\bar{r}}. \\
         \end{aligned}
    \end{equation}
    Thus we see that for different theory, the physical conjugation condition are inequivalent. \par

\subsection{Multiplet chiral super-BMS\texorpdfstring{$_4$}{4} algebra}\label{subsec:MultipletletSuperBMS4}

    In fact, there is two more self-consistent structures of super BMS$_4$ algebra, based on multiplet representations. In \cite{Chen:2023esw}, the authors introduced homogeneous and inhomogeneous super BMS$_3$ algebra. Actually, the singlet super-BMS$_4$ algebra in \eqref{eq:SingletSuperBMS4} is similar to the homogeneous case. Inspired by the inhomogeneous super-BMS$_3$ algebra, we introduce multiplet super-BMS$_4$ algebras with generators
    \begin{align}
        &\{L_n, \bar{L}_{\bar{n}}, M_{r,\bar{r}}, \mathcal{Q}^{1}_{r}, \mathcal{Q}^{2}_{m,\bar{r}}\},\label{eq:LeftSuperBMS4}\\
        &\{L_n, \bar{L}_{\bar{n}}, M_{r,\bar{r}}, \bar{\mathcal{Q}}^{1}_{\bar{r}}, \bar{\mathcal{Q}}^{2}_{r,\bar{m}}\},\label{eq:RightSuperBMS4}
    \end{align}
    where the super script $1,2$ in the supercharges label the order in the multiplets. Let us see if they lead to consistent structure. \par

    The starting point is assuming the existence of $\mathcal{Q}^1$ which is a representation of Virasoro algebra $L_n$ satisfying
    \begin{equation}\label{eq:BMS4RepsOfLn}
        [L_m,\mathcal{Q}^1_r]=\left(\frac{l}{2}m-r\right)\mathcal{Q}^1_{m+r},
    \end{equation}
    where $\left(\frac{l}{2}, 0\right)$ is the weight under $L_m$ and $\bar{L}_{\bar{m}}$. We further assume $\{\mathcal{Q}^{1}_{r},\mathcal{Q}^{1}_{s}\} \propto L_{r+s}$. Noticing the weight of $L_m$ is $(1,0)$, the Jacobi identity of $\{L_m,L_n,\mathcal{Q}^{1}_{r}\}$ then fix the weight $\frac{l}{2}=\frac{1}{2}$. The Jacobi identity of $\{L_m,\mathcal{Q}^{1}_{r},\mathcal{Q}^{1}_{s}\}$ requires the coefficient to be one, i.e., $\{\mathcal{Q}^{1}_{r},\mathcal{Q}^{1}_{s}\} = L_{r+s}$. The Jacobi identity of $\{M_{r,\bar{r}}, \mathcal{Q}^{1}_{s}, \mathcal{Q}^{1}_{t}\}$ further requires the commutation relation of $[M_{r,\bar{r}}, \mathcal{Q}^{1}_{s}]\propto \mathcal{Q}^{2}$ to be nonvanishing in order to make this algebra closed. Together with Jacobi identity of $\{L_n, M_{r,\bar{r}}, \mathcal{Q}^{1}_{s}\}$, it asks the weight of $\mathcal{Q}^{2}$ being $\left(1, \frac{1}{2}\right)$, and the commutation relations being $[M_{r,\bar{r}}, \mathcal{Q}^{1}_{s}] = \mathcal{Q}^{2}_{r+s,\bar{r}}$ and $ \{\mathcal{Q}^{1}_{r}, \mathcal{Q}^{2}_{m,\bar{s}}\} =\left(\frac{m}{2}-r\right) M_{m+r,\bar{s}}$. \par
    
    Finally, assuming that there is no bosonic generators other than $L, \bar{L}$ and $M$, the Jacobi identities of $\{M_{r,\bar{r}}, M_{s,\bar{s}}, \mathcal{Q}^{1}_t\}$, $\{M_{r,\bar{r}}, \mathcal{Q}^{1}_s, \mathcal{Q}^{2}_{m,\bar{t}}\}$ and $\{\mathcal{Q}^{1}_r, \mathcal{Q}^{2}_{m,\bar{s}}, \mathcal{Q}^{2}_{n,\bar{t}}\}$ then lead to $\{\mathcal{Q}^{2}_{m,\bar{r}}, \mathcal{Q}^{2}_{n,\bar{s}}\} = 0$, and $[M_{r,\bar{r}},\mathcal{Q}^{2}_{m,\bar{s}}]=0$. \par

    The same discussion applies to $\bar{\mathcal{Q}}^1$ with weight $\left(0, \frac{1}{2}\right)$ and $\bar{\mathcal{Q}}^2$ with weight $\left(\frac{1}{2}, 1\right)$. The Jacobi identity of $\{\mathcal{Q}^1_{r},\mathcal{Q}^1_{s},\bar{\mathcal{Q}}^1_{\bar{t}}\}$ confirms the two sets of super-generators anti-commutes $\{\mathcal{Q},\bar{\mathcal{Q}}\}=0$. While, at the same time, the Jacobi identity of $\{\mathcal{Q}^1_{r},\mathcal{Q}^2_{m,\bar{s}},\bar{\mathcal{Q}}^1_{\bar{t}}\}$ ensures that $\mathcal{Q}$ and $\bar{\mathcal{Q}}$ can not appear at the same time. \par

    Thus in conclusion, the ansatz that $\{\mathcal{Q}^{1}_{r},\mathcal{Q}^{1}_{s}\} \propto L_{r+s}$, as well as the requirement that there is no other bosonic generators than $L, \bar{L}$ and $M$, lead to an algebra containing the generators $\{L_n, \bar{L}_{\bar{n}}, M_{r,\bar{r}}, \mathcal{Q}^{1}_{r}, \mathcal{Q}^{2}_{m,\bar{r}}\}$. The commutation relations involving fermionic generators are
    \begin{equation}
        \begin{aligned}
            &[L_m, \mathcal{Q}^{1}_{r}] = \left(\frac{m}{2}-r\right)\mathcal{Q}^{1}_{m+r}, \quad 
                [M_{r,\bar{r}}, \mathcal{Q}^{1}_{s}] = \mathcal{Q}^{2}_{r+s,\bar{r}}, \\[0.5em]
            &[L_m, \mathcal{Q}^{2}_{n,\bar{r}}] = (m-n)\mathcal{Q}^{2}_{m+n,\bar{r}}, \quad
                [\bar{L}_{\bar{m}},\mathcal{Q}^{2}_{n,\bar{r}}] = \left(\frac{\bar{m}}{2}-\bar{r}\right)\mathcal{Q}^{2}_{n,\bar{m}+\bar{r}}, \\[0.5em]
            &\{\mathcal{Q}^{1}_{r},\mathcal{Q}^{1}_{s}\} = L_{r+s}, \quad
                \{\mathcal{Q}^{1}_{r},\mathcal{Q}^{2}_{m,\bar{s}}\} =\left(\frac{m}{2}-r\right) M_{m+r,\bar{s}}.\\
        \end{aligned}
    \end{equation}
    Similarly, the ansatz that $\{\bar{\mathcal{Q}}^{1}_{\bar{r}},\bar{\mathcal{Q}}^{1}_{\bar{s}}\} \propto \bar{L}_{\bar{r}+\bar{s}}$, as well as the assumption that there is no other bosonic generators than $L, \bar{L}$ and $M$, give rise to an algebra composed of the generators $\{L_n, \bar{L}_{\bar{n}}, M_{r,\bar{r}}, \bar{\mathcal{Q}}^{1}_{\bar{r}}, \bar{\mathcal{Q}}^{2}_{r,\bar{m}}\}$, with commutation relations
    \begin{equation}
        \begin{aligned}
            &[\bar{L}_{\bar{m}}, \bar{\mathcal{Q}}^{1}_{r}] = \left(\frac{\bar{m}}{2}-\bar{r}\right)\bar{\mathcal{Q}}^{1}_{\bar{m}+\bar{r}}, \quad 
                [M_{r,\bar{r}}, \bar{\mathcal{Q}}^{1}_{\bar{s}}] = \bar{\mathcal{Q}}^{2}_{r,\bar{r}+\bar{s}},\\[0.5em]
            &[\bar{L}_{\bar{m}}, \bar{\mathcal{Q}}^{2}_{r,\bar{n}}] = (\bar{m}-\bar{n})\bar{\mathcal{Q}^{2}}_{r,\bar{m}+\bar{n}}, \quad
                [L_m, \bar{\mathcal{Q}}^{2}_{r,\bar{n}}] = \left(\frac{m}{2}-r\right)\bar{\mathcal{Q}}^{2}_{m+r,\bar{n}},\\[0.5em]
            &\{\bar{\mathcal{Q}}^{1}_{\bar{r}},\bar{\mathcal{Q}}^{1}_{m,\bar{s}}\} = \bar{L}_{\bar{r}+\bar{s}}, \quad
                \{\bar{\mathcal{Q}}^{1}_{\bar{r}},\bar{\mathcal{Q}}^{2}_{s,\bar{m}}\} =\left(\frac{\bar{m}}{2}-\bar{r}\right) M_{r,\bar{m}+\bar{s}}.\\
        \end{aligned}
    \end{equation}
    Together with \eqref{eq:BMS4Algebra}, these two sets of commutation relations give two kinds of super-BMS$_4$ algebra. These two algebras are called multiplet superalgebra because under $M$, $(\mathcal{Q}^1,\mathcal{Q}^2)$ and $(\bar{\mathcal{Q}}^1,\bar{\mathcal{Q}}^2)$ are two multiplets:
    \begin{equation}
        \mathcal{Q}^1\xrightarrow{M}\mathcal{Q}^2, \quad \bar{\mathcal{Q}}^1\xrightarrow{M}\bar{\mathcal{Q}}^2.
    \end{equation}
    In fact, these algebras can be further extended with $\mathcal{Q}^{\lambda>2}$ and $\bar{\mathcal{Q}}^{\lambda>2}$. However, the algebra can only be closed after introducing more bosonic generators. Thus we find no good physical reason to investigate this possibility. Noticing that $\mathcal{Q}$ and $\bar{\mathcal{Q}}$ behave as left- and right-hand version of the supersymmetry, we can also call \eqref{eq:LeftSuperBMS4} and \eqref{eq:RightSuperBMS4} chiral super-BMS$_4$ algebras.
    \par
    
    It is important to notice that nether of the above chiral super-BMS$_4$ algebras can be found by first taking $c\to 0$ limit of a given $4$D super-Poincar\'e algebra and then searching its infinite extension. Consider the left-hand super BMS$_4$ algebra \eqref{eq:LeftSuperBMS4} for example, from the Jacobi identity of $\{M_{r,\bar{r}}, \mathcal{Q}^{1}_{s}, \mathcal{Q}^{1}_{t}\}$ we found $\mathcal{Q}^{2}$ must appears in the algebra. Thus its largest finite subalgebra is given by $m\in\{1,0,-1\}$ and $r,\bar{r}\in\left\{\frac{1}{2},-\frac{1}{2}\right\}$. In terms of Poincar\'e algebra, $\mathcal{Q}^{1}$ and $\mathcal{Q}^{2}$ are in $\left(\frac{1}{2}, 0\right)$ and $\left(1, \frac{1}{2}\right)$ representation under Lorentzian rotations $J_{\mbox{\tiny P}}\sim\{L, \bar{L}\}$ respectively, while the translations $P_{\mbox{\tiny P}}\sim M$ maps $\mathcal{Q}^{1}$ to $\mathcal{Q}^{2}$. Obviously this can not be stemmed from super-Poincar\'e algebra where the translations commutes with the supercharges. \par

    These algebras can neither be found by extending $3$D Carrollian superconformal algebra. Considering the left-hand chiral algebra for example again, the conformal dimensions of $\mathcal{Q}^{2}_{\pm 1,\pm \frac{1}{2}}$ are respectively $\Delta=\mp \frac{3}{2}$. Clearly, we should introduce fermionic generators $U$ and $V$ with $\Delta=\pm \frac{3}{2}$ in the algebra, such that the algebra is generated by 
    \begin{equation}\label{eq:LeftSuperBMS4FiniteSubAlgebra}
        \{D, J^{12}, B^i, P^\mu, K^\mu, Q^{\frac{1}{2}}_{1}, Q^{\frac{3}{2}}_{2}, Q^{-\frac{1}{2}}_{2}, S^{-\frac{1}{2}}_{1}, S^{\frac{1}{2}}_{2}, S^{-\frac{3}{2}}_{2}, U^{\frac{1}{2}}, V^{-\frac{1}{2}}\},
    \end{equation}
    where the indices of fermionic generators $F^s_\lambda$ label the spin $s$ and rank $\lambda$ for the multiplet representations under Carrollian rotations. \par

    \begin{table}[htpb]
        \renewcommand\arraystretch{1.5}
        \centering
        \caption{\centering The supercharges' conformal dimension and representations under Carrollian rotations. The arrows represent action of boost generators respectively. }
        \begin{tabular}{c|c|c|c}
            \hline
                $\Delta=\frac{3}{2}$ & $\Delta=\frac{1}{2}$ & $\Delta=-\frac{1}{2}$ & $\Delta=-\frac{3}{2}$ \\\hline
                \includegraphics[height=2.5cm]{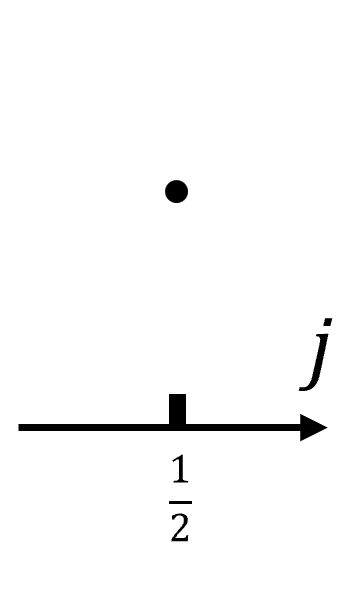} &
                \includegraphics[height=2.5cm]{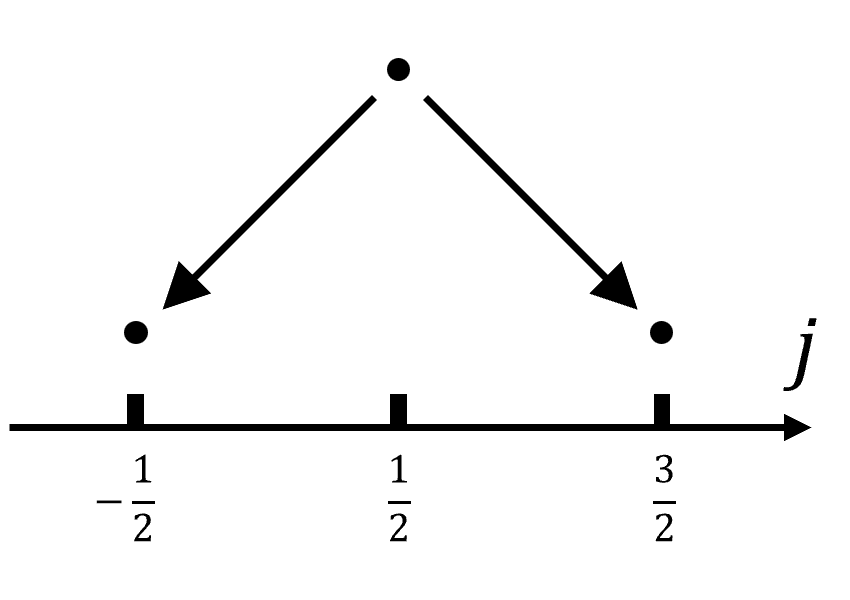} &
                \includegraphics[height=2.5cm]{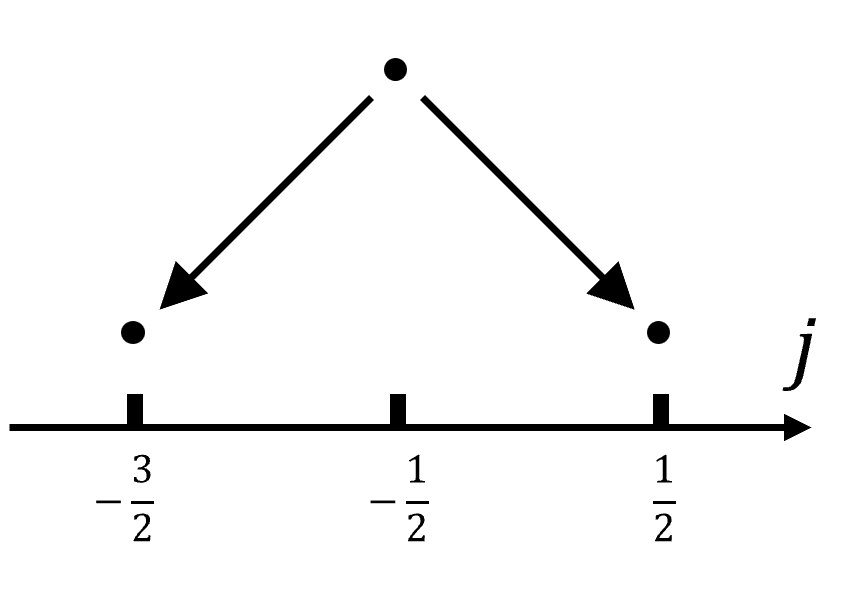} &
                \includegraphics[height=2.5cm]{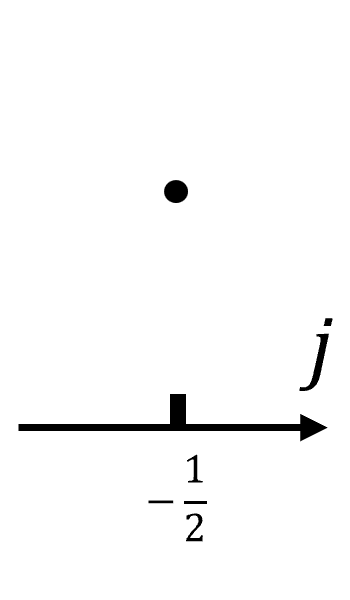} \\[-1em]
                $U$ & $Q$ & $S$ & $V$ \\\hline
        \end{tabular}
        \label{tab:MultipletSuperBMS4SuperCharges}
    \end{table}

    Now, it is time to discuss the Hermitian conjugation conditions. As discussed in Section \ref{subsec:SingletSuperBMS4}, there are six kinds of conjugation conditions on the BMS$_4$ algebra. However, here the supercharges are chiral, so only the type I conditions fit into this case, while the type II conditions which map the left-hand generators to the right-hand generators and vice versa are inconsistent with the algebra. Without losing generality, we conclude the compatible Hermitian conditions of the left-hand superalgebra in Table \ref{tab:MultipletSuperBMS4ConjugationConditions}. \par

    \begin{table}[htpb]
        \renewcommand\arraystretch{2.5}
        \centering
        \caption{\centering The Hermitian conjugation conditions of left-hand super multiplet BMS$_4$ algebra.}
        \resizebox{\linewidth}{!}{\begin{tabular}{c|c|c|c|c|c}
            \hline
                & {\Large $L_m$} & {\Large $\bar{L}_{\bar{m}}$} & {\Large $M_{r,\bar{r}}$} & {\Large $\mathcal{Q}^1_{r}$} & {\Large $\mathcal{Q}^2_{n,\bar{r}}$}\\\hline\hline
            {\large type I-11} & $- e^{i c_1 m} L_m$ & $- e^{i c_2 \bar{m}} \bar{L}_{\bar{m}}$ 
                & $e^{i c_1 r +i c_2 \bar{r}+ ic_3} M_{r,\bar{r}}$ 
                & $e^{i c_1 r + i\pi(A+\frac{1}{2})}\mathcal{Q}^1_{r}$ & $-e^{i c_1 n + i c_2 \bar{r} + i(c_3+\pi A+\frac{\pi}{2})}\mathcal{Q}^2_{n,\bar{r}}$\\\hline
            {\large type I-12} & $- e^{i c_1 m} L_m$ & $e^{c_2 \bar{m}} \bar{L}_{-\bar{m}}$ 
                & $e^{i c_1 r + c_2 \bar{r}+ ic_3} M_{r,-\bar{r}}$ 
                & $e^{i c_1 r+ i\pi(A+\frac{1}{2})}\mathcal{Q}^1_{r}$ & $-e^{i c_1 n + c_2 \bar{r} + i(c_3+\pi A+\frac{\pi}{2})}\mathcal{Q}^2_{n,-\bar{r}}$\\\hline
            {\large type I-21} & $e^{c_1 m}L_{-m}$ & $- e^{i c_2 \bar{m}} \bar{L}_{\bar{m}}$ 
                & $e^{ c_1 r +i c_2 \bar{r}+ ic_3} M_{-r,\bar{r}}$ 
                & $e^{c_1 r+ i\pi A}\mathcal{Q}^1_{-r}$ & $-e^{c_1 n + i c_2 \bar{r} + i(c_3+\pi A)}\mathcal{Q}^2_{-n,\bar{r}}$\\\hline
            {\large type I-22} & $e^{c_1 m}L_{-m}$ & $e^{c_2 \bar{m}} \bar{L}_{-\bar{m}}$ 
                & $e^{ c_1 r + c_2 \bar{r}+ ic_3} M_{-r,-\bar{r}}$ 
                & $e^{c_1 r+ i\pi A}\mathcal{Q}^1_{-r}$ & $-e^{c_1 n + c_2 \bar{r} + i(c_3+\pi A)}\mathcal{Q}^2_{-n,-\bar{r}}$\\\hline
        \end{tabular}}
        \label{tab:MultipletSuperBMS4ConjugationConditions}
    \end{table}

\section{Conclusion and Discussion}\label{sec:Discussion}

    In this paper, we conducted a detailed investigation of the supersymmetric extension of both the Carrollian algebra and the Carrollian conformal algebra in $d=4$ and $d=3$. For $4$D Carrollian algebra, we studied possible supersymmetric extensions for the supercharges transforming under singlet or chain representations of the Carrollian rotations. We saw that for specific representation patterns, super-Lie algebras exist for arbitrary spin $j$. Especially, we found two structures emerging from taking the ultra-relativistic limit of super-Poincar\'e algebra: one with $Q_1,Q_2\in(j)$, and the commutation relations shown in \eqref{eq:4dSuperCarrCase1}; the other with $Q\in (j)\to(j)$ and the commutation relations shown in \eqref{eq:4dSuperCarrCase2}. They correspond to different rescaling of the supercharges under the limit. For conformal case in $4$D, we found only three structures \eqref{eq:4dSuperConCarrSolution1}, \eqref{eq:4dSuperConCarrSolution2} and \eqref{eq:4dSuperConCarrSolution3} satisfying the basic ansatz \eqref{eq:4dsuperConCarrAnsatz}, in which only \eqref{eq:4dSuperConCarrSolution3} is nontrivial in the sense that anti-commutators of supercharges are nonvanishing. This algebra is in fact equivalent to $5$-dimensional super-Poincar\'e algebra $\mathfrak{iso}(1,4|1)$. A remarkable observation is that the Carrollian superconformal algebra achieves closure without R-symmetry generator. Instead, the possible R-symmetry is either the central charge or outer automorphisms of the algebra. \par

    In the $3$-dimensional case, the lack of constraints on the representations of Carrollian rotations is a main challenge for discussing super-Carrollian algebra. Instead of having a thorough investigation of all possible structure of 3D super-Carrollian algebra, we were satisfied focusing on the supersymmetric extension of Carrollian conformal algebra in $d=3$. We revealed that the nontrivial super-Lie algebra given by \eqref{eq:3dSuperConCarrAlg} is isomorphic to $4$-dimensional super-Poincar\'e algebra $\mathfrak{iso}(1,3|1)$. Similar to $4$D case, this algebra does not need R-symmetry to be closed. \par

    As is well established, the $3$D Carrollian conformal algebra admits an extension to BMS$_4$ algebra. In the same manner, the algebra \eqref{eq:3dSuperConCarrAlg} can also be extended to the so-called singlet super BMS$_4$ algebra, where the commutation relations involving fermionic generators are given by \eqref{eq:SingletSuperBMS4}. We also discussed various Hermitian conjugation conditions, and found that although the algebras share similar structure, the physical conjugation conditions of $4$D Poincar\'e theory, $3$D Carrollian conformal theory and $2$D Lorentzian celestial CFT are all different. Furthermore, the BMS$_4$ algebra admits two other supersymmetric extensions, i.e. the chiral multiplet super-BMS$_4$ algebras in \eqref{eq:LeftSuperBMS4} and \eqref{eq:RightSuperBMS4}. In particular, neither of them can be found by taking $c\to0$ limit. Besides, the finite dimensional subalgebras \eqref{eq:LeftSuperBMS4FiniteSubAlgebra} are not Carrollian superconformal algebra in the traditional manner because there exist $\Delta=\pm \frac{3}{2}$ fermionic generators. The multiplet chiral super BMS$_4$ algebras represent particularly noteworthy examples, as their finite-dimensional subalgebras notably differ from conventional Carrollian superconformal algebra. In subsequent work \cite{Zheng:2025rfe}, we will systematically explore additional possibilities for supersymmetric extensions of the BMS algebra and investigate the corresponding theory constructions.\par

    This work establishes important foundations for several promising avenues in supersymmetric Carrollian physics: 
    \vspace{-1.5em}
    
    \paragraph{Asymptotic supersymmetry} 
    The BMS$_4$ algebra was conventionally derived as the asymptotic symmetry algebra of flat spacetime. A natural and compelling question arises that can analogous methods be developed to derive asymptotic supersymmetry? The singlet super BMS$_4$ algebra \eqref{eq:SingletSuperBMS4} and multiplet BMS$_4$ algebras \eqref{eq:LeftSuperBMS4} and \eqref{eq:RightSuperBMS4} we found in this work serve as the targets for this investigation. The method will be significantly helpful to extend the discussions on gravity in flat background to supergravity.\par

    An important observation is that the supercharges in the left- and right-hand chiral super BMS$_4$ algebra are incompatible, unless further introducing additional bosonic generators. Crucially, these generators should not generate spacetime symmetry which has been generated by $L, \bar{L}$ and $M$ in BMS$_4$. This suggests that these extra generators must be related to gauge symmetries, which play a key role in supergravity. \par
    
    \paragraph{Construction of supersymmetric theories} 
    The algebras we established enable further constructions of explicit supersymmetric theories. For \eqref{eq:4dSuperCarrCase1} and \eqref{eq:4dSuperCarrCase2} of $4$D super-Carrollian algebra which can be derived from taking the limit, the corresponding supersymmetric theories are expected to be obtained similarly via the taking-limit procedure. However, the multiplet super BMS$_4$ algebras \eqref{eq:LeftSuperBMS4} and \eqref{eq:RightSuperBMS4} can not be obtained by taking the limit. Thus an intrinsic method is demanded to build the theories with such symmetries. \par

    Using the structure revealed in this work, we can construct various supersymmetric theories explicitly. A particularly promising example is the Carrollian super-Yang-Mills theory. It is expected to realize one of the above Carrollian superconformal symmetries. \par
    
    \paragraph{Supersymmetry in flat holography} 
   
    Introducing supersymmetry into flatspace holography represents a particularly promising research direction. The current study of celestial holography and Carrollian holography have been focusing mostly on the bosonic fields. Using the algebraic structure developed in this paper, we can systematically extend the discussions to the fermionic fields. Especially, our results lead directly to: establish holographic dictionaries for supersymmetric theories; develop bulk reconstruction for supersymmetric theories; analyze supergravity radiation in asymptotically flat spacetime. These naturally suggest to extend the analysis of soft theorems to supersymmetric contexts, which may reveal new insights into the infrared structure of supergravity theories. \par

\section*{Acknowledgments}
   We are sincerely grateful to Cihang Li for his valuable contributions to the calculations of the $4$D super-Carrollian algebra presented in this work. We are also deeply grateful to Reiko Liu and ZeZhou Hu for valuable discussions. This research was supported in part by NSFC Grant  No. 11735001, 12275004.\par
    




\bibliographystyle{JHEP}
\bibliography{refs.bib}
\end{document}